\PassOptionsToPackage{unicode}{hyperref}
\PassOptionsToPackage{hyphens}{url}
\documentclass[
  12pt,
]{article}
\usepackage{amsmath,amssymb}
\usepackage{iftex}
\ifPDFTeX
  \usepackage[T1]{fontenc}
  \usepackage[utf8]{inputenc}
  \usepackage{textcomp} 
\else 
  \usepackage{unicode-math} 
  \defaultfontfeatures{Scale=MatchLowercase}
  \defaultfontfeatures[\rmfamily]{Ligatures=TeX,Scale=1}
\fi
\usepackage{lmodern}
\ifPDFTeX\else
\fi
\IfFileExists{upquote.sty}{\usepackage{upquote}}{}
\IfFileExists{microtype.sty}{
  \usepackage[]{microtype}
  \UseMicrotypeSet[protrusion]{basicmath} 
}{}
\makeatletter
\@ifundefined{KOMAClassName}{
  \IfFileExists{parskip.sty}{%
    \usepackage{parskip}
  }{
    \setlength{\parindent}{0pt}
    \setlength{\parskip}{6pt plus 2pt minus 1pt}}
}{
  \KOMAoptions{parskip=half}}
\makeatother
\usepackage{xcolor}
\usepackage[margin=1in]{geometry}
\usepackage{color}
\usepackage{fancyvrb}

\DefineVerbatimEnvironment{Highlighting}{Verbatim}{commandchars=\\\{\}}
\usepackage{framed}
\definecolor{shadecolor}{RGB}{248,248,248}
\newenvironment{Shaded}{\begin{snugshade}}{\end{snugshade}}

\newcommand{\AttributeTok}[1]{\textcolor[rgb]{0.13,0.29,0.53}{#1}}

\newcommand{\ConstantTok}[1]{\textcolor[rgb]{0.56,0.35,0.01}{#1}}
\newcommand{\ControlFlowTok}[1]{\textcolor[rgb]{0.13,0.29,0.53}{\textbf{#1}}}

\newcommand{\DecValTok}[1]{\textcolor[rgb]{0.00,0.00,0.81}{#1}}

\newcommand{\FloatTok}[1]{\textcolor[rgb]{0.00,0.00,0.81}{#1}}
\newcommand{\FunctionTok}[1]{\textcolor[rgb]{0.13,0.29,0.53}{\textbf{#1}}}

\newcommand{\NormalTok}[1]{#1}

\newcommand{\OtherTok}[1]{\textcolor[rgb]{0.56,0.35,0.01}{#1}}

\newcommand{\SpecialCharTok}[1]{\textcolor[rgb]{0.81,0.36,0.00}{\textbf{#1}}}

\newcommand{\StringTok}[1]{\textcolor[rgb]{0.31,0.60,0.02}{#1}}

\usepackage{graphicx}
\makeatletter
\newsavebox\pandoc@box
\newcommand*\pandocbounded[1]{
  \sbox\pandoc@box{#1}%
  \Gscale@div\@tempa{\textheight}{\dimexpr\ht\pandoc@box+\dp\pandoc@box\relax}%
  \Gscale@div\@tempb{\linewidth}{\wd\pandoc@box}%
  \ifdim\@tempb\p@<\@tempa\p@\let\@tempa\@tempb\fi
  \ifdim\@tempa\p@<\p@\scalebox{\@tempa}{\usebox\pandoc@box}%
  \else\usebox{\pandoc@box}%
  \fi%
}
\def\fps@figure{htbp}
\makeatother
\NewDocumentCommand\citeproctext{}{}

\makeatletter
 \let\@cite@ofmt\@firstofone
 \def\@biblabel#1{}
 \def\@cite#1#2{{#1\if@tempswa , #2\fi}}
\makeatother
\newlength{\cslhangindent}
\newlength{\csllabelwidth}
\newenvironment{CSLReferences}[2] 
 {\begin{list}{}{%
  \setlength{\itemindent}{0pt}
  \setlength{\leftmargin}{0pt}
  \setlength{\parsep}{0pt}
  \ifodd #1
   \setlength{\leftmargin}{\cslhangindent}
   \setlength{\itemindent}{-1\cslhangindent}
  \fi
  \setlength{\itemsep}{#2\baselineskip}}}
 {\end{list}}
\usepackage{calc}

\usepackage{bookmark}
\IfFileExists{xurl.sty}{\usepackage{xurl}}{} 
\hypersetup{
  pdftitle={Global Minima by Penalized Full-dimensional Scaling},
  pdfauthor={Jan de Leeuw},
  hidelinks,
  pdfcreator={LaTeX via pandoc}}

\title{Global Minima by Penalized Full-dimensional Scaling}
\author{Jan de Leeuw}
\date{First created on May 06, 2019. Last update on July 23, 2024}

\begin{document}
\maketitle
\begin{abstract}
The full-dimensional (metric, Euclidean, least squares) multidimensional
scaling stress loss function is combined with a quadratic external
penalty function term. The trajectory of minimizers of stress for
increasing values of the penalty parameter is then used to find
(tentative) global minima for low-dimensional multidimensional scaling.
This is illustrated with several one-dimensional and two-dimensional
examples.
\end{abstract}

{
\setcounter{tocdepth}{3}
\tableofcontents
}
\textbf{Note:} This is a working paper which will be expanded/updated
frequently. All suggestions for improvement are welcome. The directory
\href{https://www.github.com/deleeuw/penalty}{github.com/deleeuw/penalty}
has a pdf version, a html version, the bib files, the complete Rmd file
with the code chunks, and the R source code.

\section{Introduction}\label{introduction}

Full-dimensional Scaling (FDS) was introduced by De Leeuw (1993). De
Leeuw, Groenen, and Mair (2016) discuss it in some detail. In FDS we
minimize the usual Multidimensional Scaling (MDS) least squares loss
function first used by Kruskal (1964a) and Kruskal (1964b).
\begin{equation}\label{E:stress}
\sigma(Z)=\frac12\mathop{\sum\sum}_{1\leq i<j\leq n}w_{ij}(\delta_{ij}-d_{ij}(Z))^2
\end{equation} over all \(n\times n\) \emph{configuration matrices}
\(Z\). The loss at \(Z\) is often called the \emph{stress} of
configuration \(Z\). More generally we define pMDS as the problem of
minimizing \(\eqref{E:stress}\) over all \(n\times p\) matrices. Thus
FDS is the same as nMDS. If a configuration \(Z\) has \(n\) columns
(i.e.~is square) it is called a \emph{full configuration}.

In \(\eqref{E:stress}\) the matrices \(W=\{w_{ij}\}\) and
\(\Delta=\{\delta_{ij}\}\) of \emph{weights} and \emph{dissimilarities}
are non-negative, symmetric, and hollow. To simplify matters we suppose
both \(W\) and \(\Delta\) have positive off-diagonal elements. The
matrix \(D(X)=\{d_{ij}(Z)\}\) has the \emph{Euclidean distances} between
the rows of the configuration \(Z\). Thus \[
d_{ij}(Z)=\sqrt{(z_i-z_j)'(z_i-z_j)}.
\] We now introduce some standard MDS notation, following De Leeuw
(1977). Define the matrix \(V=\{v_{ij}\}\) by
\begin{equation}\label{E:V}
v_{ij}=\begin{cases}-w_{ij}&\text{ if }i\not= j,\\\sum_{j=1}^n w_{ij}&\text{ if }i=j,\end{cases}
\end{equation} and the matrix valued function \(B(Z)=\{b_{ij}(Z)\}\) by
\begin{equation}\label{E:B}
b_{ij}(Z)=\begin{cases}-w_{ij}e_{ij}(Z)&\text{ if }i\not= j,\\\sum_{j=1}^n w_{ij}e_{ij}(Z)&\text{ if }i=j,\end{cases}
\end{equation} where \(E(Z)=\{e_{ij}(Z)\}\) is defined as
\begin{equation}\label{E:E}
e_{ij}(Z)=\begin{cases}\frac{\delta_{ij}}{d_{ij}(Z)}&\text{ if }d_{ij}(Z)>0,\\0&\text{ otherwise}.\end{cases}
\end{equation}

Note that \(V\) and \(B(Z)\) are both positive semi-definite and
doubly-centered. Matrix \(V\) has rank \(n-1\). If all off-diagonal
\(d_{ij}(Z)\) are positive then \(B(Z)\) has rank \(n-1\) for all \(Z\).
Note that De Leeuw (1984) established that near a local minimum of
stress all off-diagonal distances are indeed positive. The only vectors
in the null-space of both \(V\) and \(B(Z)\) are the vectors
proportional to the vector with all elements equal to one.

We assume in addition, without loss of generality, that \[
\frac12\mathop{\sum\sum}_{1\leq i<j\leq n}w_{ij}\delta_{ij}^2=1.
\] With these definitions we can rewrite the stress \(\eqref{E:stress}\)
as \begin{equation}\label{E:mstress}
\sigma(Z)=1-\mathbf{tr}\ Z'B(Z)Z+\frac12\mathbf{tr}\ Z'VZ,
\end{equation} and we can write the stationary equations as
\begin{equation}\label{E:stationary}
(V-B(Z))Z=0,
\end{equation} or, in fixed point form, \(Z=V^+B(Z)Z\).

Equation \(\eqref{E:mstress}\) shows, by the way, something which is
already obvious from \(\eqref{E:stress}\). Distances are invariant under
translation. This is reflected in \(B(Z)\) and \(V\) being
doubly-centered. As a consequence we usually require, again without loss
of generality, that \(Z\) is column-centered. And that implies that
\(Z\) has rank at most \(n-1\), which means that FDS is equivalent to
minimizing stress over all \(n\times (n-1)\) matrices, which we can
assume to be column-centered as well. Configurations with \(n-1\)
columns can be called full configurations as well. In addition,
distances are invariant under rotation, and consequently if \(Z\) solves
the stationary equations with value \(\sigma(Z)\) then \(ZK\) solves the
stationary equations for all rotation matrices \(K\), and
\(\sigma(ZK)=\sigma(K)\). This means there are no isolated local minima
in configuration space, each local minimum is actually a continuum of
rotated matrices in \(\mathbb{R}^{n\times n}\). This is a nuisance in
the analysis of FDS and pMDS that is best dealt with by switching to the
parametrization outlined in De Leeuw (1993).

\section{Convex FDS}\label{convex-fds}

Instead of defining the loss function \(\eqref{E:stress}\) on the space
of all \(n\times n\) configuration matrices \(Z\) we can also define it
over the space of all positive semidefinite matrices \(C\) of order
\(n\). This gives \begin{equation}\label{E:cmds}
\sigma(C)=1-\mathop{\sum\sum}_{1\leq i<j\leq n}w_{ij}\delta_{ij}\sqrt{c_{ii}+c_{jj}-2c_{ij}}+\frac12\mathbf{tr}\ VC.
\end{equation} The Convex Full-dimensional Scaling (CFDS) problem is to
minimize loss function \(\eqref{E:cmds}\) over all \(C\gtrsim 0\).
Obviously if \(Z\) minimizes \(\eqref{E:stress}\) then \(C=ZZ'\)
minimizes \(\eqref{E:cmds}\). And, conversely, if \(C\) minimizes
\(\eqref{E:cmds}\) then any \(Z\) such that \(C=ZZ'\) minimizes
\(\eqref{E:stress}\).

The definition \(\eqref{E:cmds}\) shows that the CFDS loss function is a
convex function on the cone of positive semi-definite matrices, because
the square root of a non-negative linear function of the elements of
\(C\) is concave. Positivity of the weights and dissimilarities implies
that loss is actually strictly convex. The necessary and sufficient
conditions for \(C\) to be the unique solution of the CFDS problem are
simply the conditions for a proper convex function to attain its minimum
at \(C\) on a closed convex cone (Rockafellar (1970), theorem 31.4).
\begin{align*}
V-B(C)&\gtrsim 0,\\
C&\gtrsim 0,\\
\mathbf{tr}\ C(V-B(C))&=0.
\end{align*} The conditions say that \(C\) and \(V-B(C)\) must be
positive semi-definite and have complimentary null spaces.

By the same reasoning as in the full configuration case, we also see
that CFDS is equivalent to maximizing \(\eqref{E:cmds}\) over all
doubly-centered positive semi-definite matrices.

If \(C\) is the solution of the CFDS problem then \(\mathbf{rank}(C)\)
is called the \emph{Gower rank} of the MDS problem defined by \(W\) and
\(\Delta\) (De Leeuw (2016)). Although there is a unique Gower rank
associated with each CFDS problem, we can also talk about the
\emph{approximate Gower rank} by ignoring the small eigenvalues of
\(C\).

\section{FDS using SMACOF}\label{fds-using-smacof}

The usual SMACOF algorithm can be applied to FDS as well. The iterations
start with \(Z^{(0)}\) and use the update rule
\begin{equation}\label{E:smacof}
Z^{(k+1)}=V^+B(Z^{(k)})Z^{(k)},
\end{equation} where \(V^+\) is the Moore-Penrose inverse of \(V\), and
is consequently also doubly-centered. This means that all \(Z^{(k)}\) in
the SMACOF sequence, except possibly \(Z^{(0)}\), are column-centered
and of rank at most \(n-1\). Equation \(\eqref{E:smacof}\) also shows
that if \(Z^{(0)}\) is of rank \(p<n-1\) then all \(Z^{(k)}\) are of
rank \(p\) as well.

De Leeuw (1977) shows global convergence of the SMACOF sequence for
pMDS, generated by \(\eqref{E:smacof}\), to a stationary point, i.e.~a
point satisfying \((V-B(Z))Z=0\). This result also applies, of course,
to nMDS, i.e.~FDS. If \(Z\) is a solution of the stationary equations
then with \(C=ZZ'\) we have both \((V-B(C))C=0\) and \(C\gtrsim 0\), but
since we generally do not have \(V-B(Z)\gtrsim 0\), this does not mean
that \(C\) solves the CFDS problem.

In fact, suppose the unique CMDS solution has Gower rank \(r\geq 2\).
Start the SMACOF FDS iterations \(\eqref{E:smacof}\) with \(Z^{(0)}\) of
the form \(Z^{(0)}=\begin{bmatrix}X^{(0)}&\mid&0\end{bmatrix}\), where
\(X^{(0)}\) is an \(n\times p\) matrix of rank \(p<r\). All \(Z^{(k)}\)
will be of this form and will also be of rank \(p\), and all
accumulation points \(Z\) of the SMACOF sequence will have this form and
\(\mathbf{rank}(Z)\leq p\). Thus \(C=ZZ'\) cannot be the solution of the
CMDS problem.

The next result shows that things are allright, after all. Although
stress in FDS is certainly not a convex function of \(Z\), it remains
true that all local minima are global.

\textbf{Lemma 1: {[}Expand{]}} If FDS stress has a local minimum at
\(\begin{bmatrix}X&\mid&0\end{bmatrix}\), where \(X\) is \(n\times p\)
and the zero block is \(n\times q\) with \(q>1\), then

1: \(\mathcal{D}\sigma(X)=(V-B(X))X=0\).

2: \(\mathcal{D}^2\sigma(X)\gtrsim 0\).

3: \(V-B(X)\gtrsim 0\).

\textbf{Proof:} We use the fact that stress is differentiable at a local
minimum (De Leeuw (1984)). If
\(Z=\begin{bmatrix}X&\mid&0\end{bmatrix}+\epsilon\begin{bmatrix}P&\mid&Q\end{bmatrix}\)
then we must have \(\sigma(Z)\geq\sigma(X)\) for all \(P\) and \(Q\).
Now \begin{multline}\label{E:expand}
\sigma(Z)=\sigma(X)+\epsilon\ \text{tr}\ P'\mathcal{D}\sigma(X)\ +\\+\frac12\epsilon^2\ \mathcal{D}^2\sigma(X)(P,P)+\frac12\epsilon^2\ \text{tr}\ Q'(V-B(X))Q+o(\epsilon^2).
\end{multline} The lemma follows from this expansion. \(\blacksquare\)

\textbf{Theorem 1: {[}FDS Local Minima{]}} If stationary point \(Z\) of
FDS is a local minimum, then it also is the global minimum, and
\(C=ZZ'\) solves the CFDS problem.

\textbf{Proof:} We start with a special case. Suppose \(Z\) is a
doubly-centered solution of the FDS stationary equations with
\(\mathbf{rank}(Z)=n-1\). Then \((V-B(Z))Z=0\) implies \(V=B(Z)\), which
implies \(\delta_{ij}=d_{ij}(Z)\) for all \(i,j\). Thus \(\sigma(Z)=0\),
which obviously is the global minimum.

Now suppose \(Z\) is a doubly-centered local minimum solution of the FDS
stationary equations with \(\mathbf{rank}(Z)=r<n-1\). Without loss of
generality we assume \(Z\) is of the form
\(Z=\begin{bmatrix}X&\mid&0\end{bmatrix}\), with \(X\) an \(n\times r\)
matrix of rank \(r\). For \(C=ZZ'\) to be a solution of the CFDS problem
it is necessary and sufficient that \(V-B(Z)\gtrsim 0\). Lemma 1 shows
that this is indeed the case at a local minimum. \(\blacksquare\)

\textbf{Corrollary 1: {[}Saddle{]}} A pMDS solution of the stationary
equations with \(Z\) singular is a saddle point.

\textbf{Corrollary 2: {[}Nested{]}} Solutions of the stationary
equations of pMDS are saddle points of qMDS with \(q>p\).

The proof of lemma 1 shows that for any \(n\times p\) configuration
\(Z\), not just for solutions of the FDS stationary equations, if
\(V-B(Z)\) is indefinite we can decrease loss by adding another
dimension. If \(Z\) is a stationary point and \(V-B(Z)\) is positive
semi-definite then we actually have found the CFDS solution, the Gower
rank, and the global minimum (De Leeuw (2014)).

\section{Penalizing Dimensions}\label{penalizing-dimensions}

In Shepard (1962a) and Shepard (1962b) a nonmetric multidimensional
scaling technique is developed which minimizes a loss function over
configurations in full dimensionality \(n-1\). In that sense the
technique is similar to FDS. Shepard's iterative process aims to
maintain monotonicity between distances and dissimilarities and at the
same time concentrate as much of the variation as possible in a small
number of dimensions (De Leeuw (2017)).

Let us explore the idea of concentrating variation in \(p<n-1\)
dimensions, but use an approach which is quite different from the one
used by Shepard. We remain in the FDS framework, but we aim for
solutions in \(p<n-1\) dimensions by penalizing \(n-p\) dimensions of
the full configuration, using the classical Courant quadratic penalty
function.

Partition a full configuration
\(Z=\begin{bmatrix}X&\mid&Y\end{bmatrix}\), with \(X\) of dimension
\(n\times p\) and \(Y\) of dimension \(n\times(n-p)\). Then
\begin{equation}\label{E:part}   
\sigma(Z)=1-\mathbf{tr}\ X'B(Z)X - \mathbf{tr}\ Y'B(Z)Y+\frac12 \mathbf{tr}\ X'VX+\frac12 \mathbf{tr}\ Y'VY.
\end{equation} Also define the \emph{penalty term}
\begin{equation}\label{E:tau}
\tau(Y)=\frac12\mathbf{tr}\ Y'VY,
\end{equation} and \emph{penalized stress} \begin{equation}\label{E:pi}
\pi(Z,\lambda)=\sigma(Z)+\lambda\ \tau(Y).
\end{equation}

Our proposed method is to minimize penalized stress over \(Z\) for a
sequence of values \(0=\lambda_1<\lambda_2<\cdots\lambda_m\). For
\(\lambda=0\) this is simply the FDS problem, for which we know we can
compute the global minimum. For fixed \(0<\lambda<+\infty\) this is a
Penalized FDS or PFDS problem. PFDS problems with increasing values of
\(\lambda\) generate a \emph{trajectory} \(Z(\lambda)\) in configuration
space.

The general theory of exterior penalty functions, which we review in
appendix A of this paper, shows that increasing \(\lambda\) leads to an
increasing sequence of stress values \(\sigma\) and a decreasing
sequence of penalty terms \(\tau\). If \(\lambda\rightarrow+\infty\) we
approximate the global minimum of the FDS problem with \(Z\) of the form
\(Z=\begin{bmatrix}X&\mid&0\end{bmatrix}\), i.e.~of the pMDS problem.
This assumes we do actually compute the global minimum for each value of
\(\lambda\), which we hope we can do because we start at the FDS global
minimum, and we slowly increase \(\lambda\). There is also a local
version of the exterior penalty result, which implies that
\(\lambda\rightarrow\infty\) takes us to a local minimum of pMDS, so
there is always the possibility of taking the wrong trajectory to a
local minimum of pMDS.

\subsection{Local Minima}\label{local-minima}

The stationary equations of the PFDS problem are solutions to the
equations \begin{align}
(V-B(Z))X&=0,\\
((1+\lambda)V-B(Z))Y&=0.
\end{align}

We can easily related stationary points and local minima of the FDS and
PFDS problem.

\textbf{Theorem 2: {[}PFDS Local Minima{]}}

1: If \(X\) is a stationary point of the pMDS problem then
\(Z=[X\mid 0]\) is a stationary point of the PFDS problem, no matter
what \(\lambda\) is.

2: If \(Z=[X\mid 0]\) is a local minimum of the PFDS problem then \(X\)
is a local minimum of pMDS and \((1+\lambda)V-B(X)\gtrsim 0\), or
\(\lambda\geq\|V^+B(X)\|_\infty-1\), with \(\|\bullet\|_\infty\) the
spectral radius (largest eigenvalue).

\textbf{Proof:}

Part 1 follows by simple substitution in the stationary equations.

Part 2 follows from the expansion for
\(Z=[X+\epsilon P\mid\epsilon Q]\). \begin{multline}\label{E:expand2}
\pi(Z)=\pi(X)+\epsilon\ \text{tr}\ P'\mathcal{D}\sigma(X)\ +\\+\frac12\epsilon^2\ \mathcal{D}^2\sigma(X)(P,P)+\frac12\epsilon^2\ \text{tr}\ Q'((1+\lambda)V-B(X))Q+o(\epsilon^2).
\end{multline} At a local minimum we must have
\(\mathcal{D}\sigma(X)=0\) and \(\mathcal{D}^2\sigma(X)(P,P)\gtrsim 0\),
which are the necessary conditions for a local minimum of pMDS. We also
must have \(((1+\lambda)V-B(X))\gtrsim 0\). \(\blacksquare\)

Note that the conditions in part 2 of theorem 2 are also sufficient for
PFDS to have a local minimum at \([X\mid 0]\), provided we eliminate
translational and rotational indeterminacy by a suitable
reparametrization, as in De Leeuw (1993).

\section{Algorithm}\label{algorithm}

The SMACOF algorithm for penalized stress is a small modification of the
unpenalized FDS algorithm \(\eqref{E:smacof}\). We start our iterations
for \(\lambda_j\) with the solution for \(\lambda_{j-1}\) (the starting
solution for \(\lambda_1=0\) can be completely arbitrary). The update
rules for fixed \(\lambda\) are

\begin{align}
X^{(k+1)}&=V^+B(Z^{(k)})X^{(k)},\\
Y^{(k+1)}&=\frac{1}{1+\lambda}V^+B(Z^{(k)})Y^{(k)}.
\end{align}

Thus we compute the FDS update \(Z^{(k+1)}=V^+B(Z^{(k)})Z^{(k)}\) and
then divide the last \(n-p\) columns by \(1+\lambda\).

Code is in the appendix. Let us analyze a number of examples.

\section{Examples}\label{examples}

This section has a number of two-dimensional and a number of
one-dimensional examples. The one-dimensional examples are of interest,
because of the documented large number of local minima of stress in the
one-dimensional case, and the fact that for small and medium \(n\) exact
solutions are available (for example, De Leeuw (2005)). By default we
use \texttt{seq(0,\ 1,\ length\ =\ 101)} for \(\lambda\) in most
examples, but for some of them we dig a bit deeper and use longer
sequences with smaller increments.

If for some value of \(\lambda\) the penalty term drops below the small
cutoff \(\gamma\), for example \ensuremath{10^{-10}}, then there is not
need to try larger values of \(\lambda\), because they will just repeat
the same result. We hope that result is the global minimum of the 2MDS
problem.

The output for each example is a table in which we give, the minimum
value of stress, the value of the penalty term at the minimum, the value
of \(\lambda\), and the number of iterations needed for convergence.
Typically we print for the first three, the last three, and some
regularly spaced intermediate values of \(\lambda\). Remember that the
stress values increase with increasing \(\lambda\), and the penalty
values decrease.

For two-dimensional examples we plot all two-dimensional configurations,
after rotating to optimum match (using the function \texttt{matchMe()}
from the appendix). We connect corresponding points for different values
of \(\lambda\). Points corresponding to the highest value of \(\lambda\)
are labeled and have a different plot symbol. For one-dimensional
examples we put \texttt{1:n} on the horizontal axes and plot the single
dimension on the vertical axis, again connecting corresponding points.
We label the points corresponding with the highest value of \(\lambda\),
and draw horizontal lines through them to more clearly show their order
on the dimension.

The appendix also has code for the function \texttt{checkUni()}, which
we have used to check the solutions in the one dimensional case are
indeed local minima. The function checks the necessary condition for a
local minimum \(x=V^+u\), with \[
u_i=\sum_{j=1}^nw_{ij}\delta_{ij}\ \mathbf{sign}\ (x_i-x_j).
\] It should be emphasized that all examples are just meant to study
convergence of penalized FDS. There is no interpretation of the MDS
results

\subsection{Chi Squares}\label{chi-squares}

In this example, of order 10, the \(\delta_{ij}\) are independent draws
from a chi-square distribution with two degrees of freedom. There is no
structure in this example, everything is random.

\begin{verbatim}
## itel  198 lambda   0.000000 stress 0.175144 penalty 0.321138 
## itel    5 lambda   0.010000 stress 0.175156 penalty 0.027580 
## itel    3 lambda   0.020000 stress 0.175187 penalty 0.025895 
## itel    1 lambda   0.100000 stress 0.175914 penalty 0.015172 
## itel    1 lambda   0.200000 stress 0.177666 penalty 0.004941 
## itel    4 lambda   0.300000 stress 0.178912 penalty 0.000088 
## itel    6 lambda   0.310000 stress 0.178933 penalty 0.000020 
## itel   20 lambda   0.320000 stress 0.178939 penalty 0.000000
\end{verbatim}

\begin{center}\includegraphics{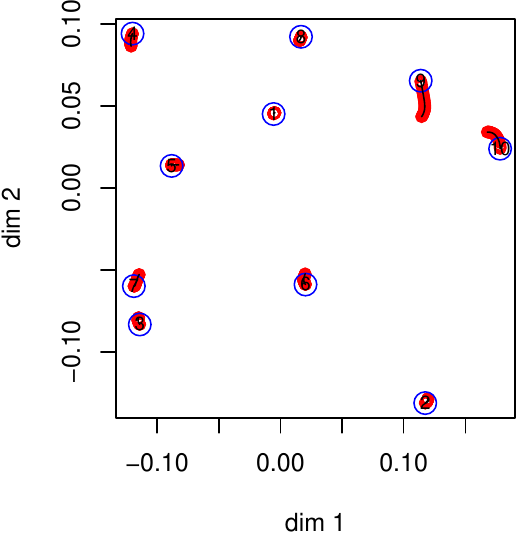} \end{center}

It seems that in this example the first two dimensions of FDS are
already close to optimal for 2MDS. This is because the Gower rank of the
dissimilarities is only three (or maybe four, the fourth singular value
of the FDS solution \(Z\) is very small).

\subsection{Regular Simplex}\label{regular-simplex}

The regular simplex has all dissimilarities equal to one. We use an
example with \(n=10\), for which the global minimum (as far as we know)
of pMDS with \(p=2\) is a configuration with nine points equally spaced
on a circle and one point in the center.

\begin{verbatim}
## itel    1 lambda   0.000000 stress 0.000000 penalty 0.400000 
## itel    7 lambda   0.010000 stress 0.000103 penalty 0.375240 
## itel    5 lambda   0.020000 stress 0.000427 penalty 0.360212 
## itel    2 lambda   0.100000 stress 0.009438 penalty 0.255499 
## itel    1 lambda   0.200000 stress 0.031826 penalty 0.152804 
## itel    1 lambda   0.300000 stress 0.059136 penalty 0.079739 
## itel    1 lambda   0.400000 stress 0.081607 penalty 0.038004 
## itel    1 lambda   0.500000 stress 0.097400 penalty 0.015499 
## itel    1 lambda   0.600000 stress 0.106700 penalty 0.004060 
## itel    1 lambda   0.700000 stress 0.109679 penalty 0.000427 
## itel    1 lambda   0.710000 stress 0.109756 penalty 0.000320 
## itel   92 lambda   0.720000 stress 0.109880 penalty 0.000000
\end{verbatim}

\begin{center}\includegraphics{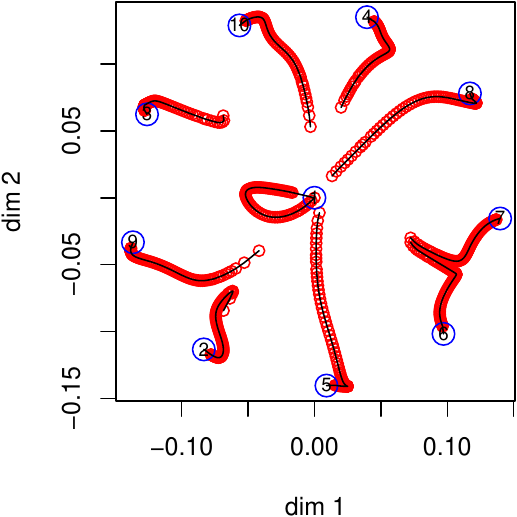} \end{center}

Next, we look at the regular simplex with \(n=4\), for which the global
minimum has four points equally spaced on a circle (i.e.~in the corners
of a square). We use \texttt{seq(0,\ 1,\ length\ =\ 101)} for the
\(\lambda\) sequence.

\begin{verbatim}
## itel    1 lambda   0.000000 stress 0.000000 penalty 0.250000 
## itel    2 lambda   0.010000 stress 0.000033 penalty 0.162166 
## itel    2 lambda   0.020000 stress 0.000158 penalty 0.156683 
## itel    2 lambda   0.100000 stress 0.004569 penalty 0.103311 
## itel    1 lambda   0.200000 stress 0.016120 penalty 0.041571 
## itel    1 lambda   0.300000 stress 0.026039 penalty 0.007635 
## itel    1 lambda   0.330000 stress 0.027355 penalty 0.003658 
## itel    5 lambda   0.340000 stress 0.028272 penalty 0.000944 
## itel   36 lambda   0.350000 stress 0.028595 penalty 0.000000
\end{verbatim}

\begin{center}\includegraphics{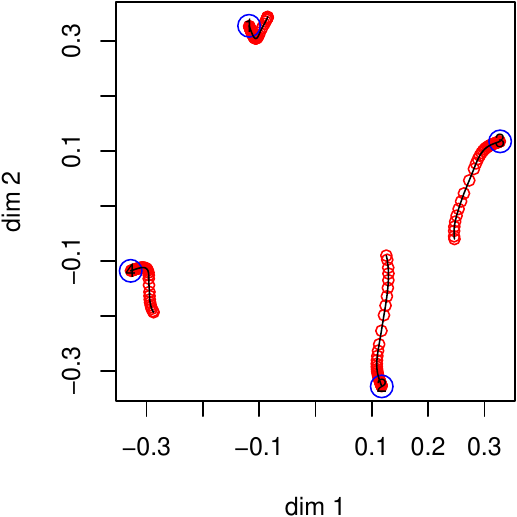} \end{center}

The solution converges to an equilateral triangle with the fourth point
in the centroid. This is a local minimum. What basically happens is that
the first two dimensions of the FDS solution are too close to the local
minimum. Or, what amounts to the same thing, the Gower rank is too large
(it is \(n-1\) for a regular simplex) , there is too much variation in
the higher dimensions, and as a consequence the first two dimensions of
FDS are a bad 2MDS solution. We try to repair this by refining the
trajectory, using \texttt{seq(0,\ 1,\ 10001)}.

\begin{verbatim}
## itel    1 lambda   0.000000 stress 0.000000 penalty 0.250000 
## itel    2 lambda   0.000100 stress 0.000000 penalty 0.166621 
## itel    2 lambda   0.000200 stress 0.000000 penalty 0.166563 
## itel    1 lambda   0.202200 stress 0.028595 penalty 0.000001 
## itel    1 lambda   0.202300 stress 0.028595 penalty 0.000001 
## itel    1 lambda   0.202400 stress 0.028595 penalty 0.000001
\end{verbatim}

\begin{center}\includegraphics{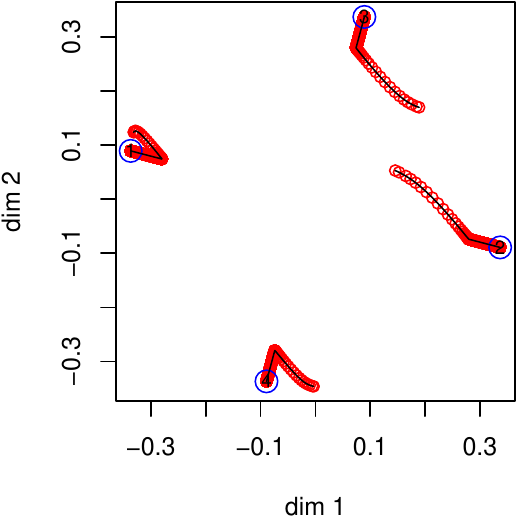} \end{center}

Now the trajectories move us from what starts out similar to an
equilateral triangle to the corners of the square, and thus we do find
the global minimum in this way. It is remarkable that we manage to find
the square even when we start closer to the triangle with midpoint.

\subsection{Intelligence}\label{intelligence}

These are correlations between eight intelligence tests, taken from the
\texttt{smacof} package. We convert to dissimilarities by taking the
negative logarithm of the correlations. As in the chi-square example,
the FDS and the 2MDS solution are very similar and the PMDS trajectories
are short.

\begin{verbatim}
## itel 2951 lambda   0.000000 stress 0.107184 penalty 7.988384 
## itel    7 lambda   0.010000 stress 0.107560 penalty 0.685654 
## itel    4 lambda   0.020000 stress 0.108528 penalty 0.628538 
## itel    3 lambda   0.030000 stress 0.110045 penalty 0.573208 
## itel    3 lambda   0.040000 stress 0.112449 penalty 0.510730 
## itel    2 lambda   0.050000 stress 0.114714 penalty 0.464650 
## itel    2 lambda   0.060000 stress 0.117623 penalty 0.415037 
## itel    2 lambda   0.070000 stress 0.121095 penalty 0.364536 
## itel    2 lambda   0.080000 stress 0.125010 penalty 0.315023 
## itel    2 lambda   0.090000 stress 0.129226 penalty 0.267831 
## itel    2 lambda   0.100000 stress 0.133589 penalty 0.223898 
## itel    2 lambda   0.110000 stress 0.137944 penalty 0.183868 
## itel    3 lambda   0.120000 stress 0.143921 penalty 0.133739 
## itel    2 lambda   0.130000 stress 0.147473 penalty 0.106166 
## itel    4 lambda   0.140000 stress 0.153215 penalty 0.064499 
## itel    4 lambda   0.150000 stress 0.157159 penalty 0.037735 
## itel    9 lambda   0.160000 stress 0.161434 penalty 0.010337 
## itel   72 lambda   0.170000 stress 0.163122 penalty 0.000000
\end{verbatim}

\begin{center}\includegraphics{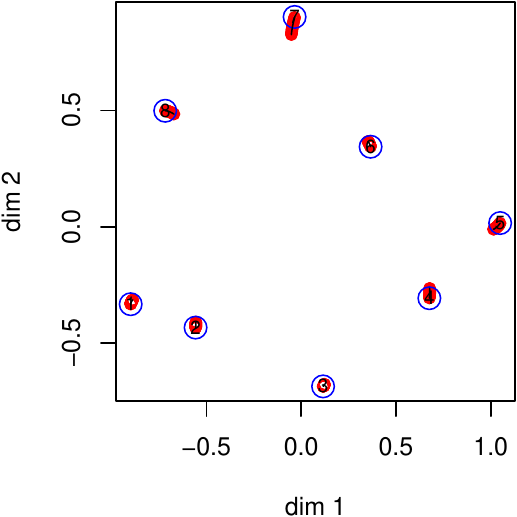} \end{center}

The singular values of the FDS solution are 1.78e+00, 1.36e+00,
5.94e-01, 1.47e-01, 3.29e-03, 1.87e-16, 4.37e-17, 3.78e-18, which shows
that the Gower rank is probably five, but approximately two.

\subsection{Countries}\label{countries}

This is the \texttt{wish} dataset from the 'smacof` package, with
similarities between 12 countries. They are converted to dissimilarties
by subtracting each of them from seven.

\begin{verbatim}
## itel 1381 lambda   0.000000 stress 4.290534 penalty 98.617909 
## itel    4 lambda   0.010000 stress 4.301341 penalty 36.137074 
## itel    3 lambda   0.020000 stress 4.336243 penalty 34.389851 
## itel    1 lambda   0.100000 stress 5.187917 penalty 23.300775 
## itel    1 lambda   0.200000 stress 7.539228 penalty 11.543635 
## itel    1 lambda   0.300000 stress 9.901995 penalty 4.963372 
## itel    1 lambda   0.400000 stress 11.523357 penalty 1.859569 
## itel    1 lambda   0.500000 stress 12.391692 penalty 0.556411 
## itel    1 lambda   0.590000 stress 12.696493 penalty 0.080144 
## itel    1 lambda   0.600000 stress 12.708627 penalty 0.060113 
## itel  100 lambda   0.610000 stress 12.738355 penalty 0.000000
\end{verbatim}

\begin{center}\includegraphics{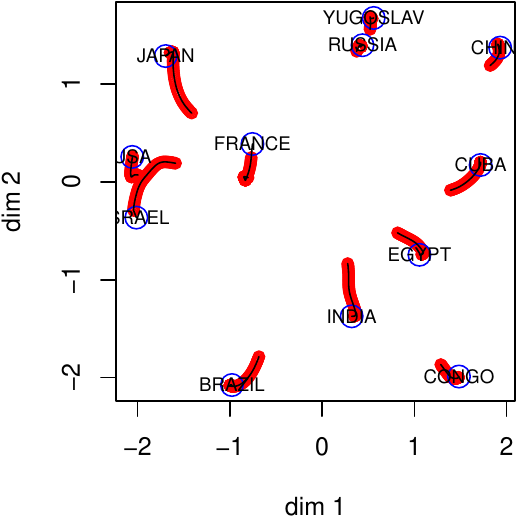} \end{center}

The singular values of the FDS solution are 4.20e+00, 3.71e+00,
2.67e+00, 1.80e+00, 1.33e+00, 6.64e-01, 5.97e-04, 7.74e-16, 3.18e-16,
2.25e-16, 8.42e-17, 4.53e-18, and the Gower rank is six or seven.

\subsection{Dutch Political Parties}\label{dutch-political-parties}

In 1967 one hundred psychology students at Leiden University judged the
similarity of nine Dutch political parties, using the complete method of
triads (De Gruijter (1967)). Data were aggregated and converted to
dissimilarities. We first print the matrix of dissimilarities.

\begin{verbatim}
##      KVP   PvdA  VVD   ARP   CHU   CPN   PSP   BP    D66  
## KVP  0.000 0.209 0.196 0.171 0.179 0.281 0.250 0.267 0.230
## PvdA 0.209 0.000 0.250 0.210 0.231 0.190 0.171 0.269 0.204
## VVD  0.196 0.250 0.000 0.203 0.185 0.302 0.281 0.257 0.174
## ARP  0.171 0.210 0.203 0.000 0.119 0.292 0.250 0.271 0.228
## CHU  0.179 0.231 0.185 0.119 0.000 0.290 0.263 0.259 0.225
## CPN  0.281 0.190 0.302 0.292 0.290 0.000 0.152 0.236 0.276
## PSP  0.250 0.171 0.281 0.250 0.263 0.152 0.000 0.256 0.237
## BP   0.267 0.269 0.257 0.271 0.259 0.236 0.256 0.000 0.274
## D66  0.230 0.204 0.174 0.228 0.225 0.276 0.237 0.274 0.000
\end{verbatim}

The trajectories from FDS to 2MDS show some clear movement, especially
of the D'66 party, which was new at the time.

\begin{verbatim}
## itel  223 lambda   0.000000 stress 0.000000 penalty 0.414526 
## itel    5 lambda   0.010000 stress 0.000061 penalty 0.196788 
## itel    2 lambda   0.020000 stress 0.000199 penalty 0.190472 
## itel    1 lambda   0.100000 stress 0.004399 penalty 0.136576 
## itel    1 lambda   0.200000 stress 0.015811 penalty 0.075466 
## itel    1 lambda   0.300000 stress 0.028235 penalty 0.036636 
## itel    1 lambda   0.400000 stress 0.038275 penalty 0.012608 
## itel    1 lambda   0.500000 stress 0.043644 penalty 0.002156 
## itel    1 lambda   0.520000 stress 0.044091 penalty 0.001324 
## itel    1 lambda   0.530000 stress 0.044253 penalty 0.001019 
## itel  277 lambda   0.540000 stress 0.044603 penalty 0.000000
\end{verbatim}

\begin{center}\includegraphics{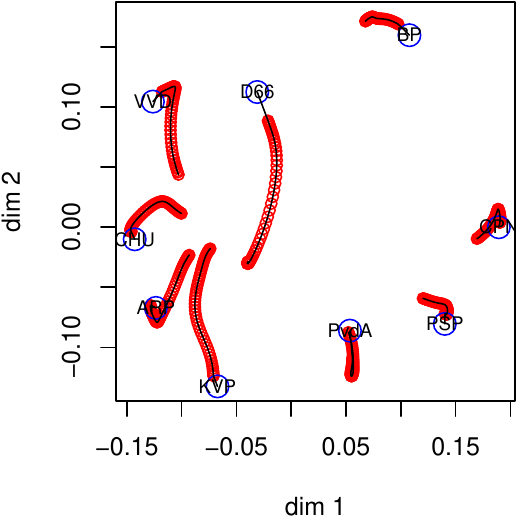} \end{center}

There seems to be some bifurcation going on at the end, so we repeat the
analysis using \texttt{seq(0,\ 1,\ length\ =\ 1001)} for \(\lambda\).
The results turn out to be basically the same.

\begin{verbatim}
## itel  223 lambda   0.000000 stress 0.000000 penalty 0.414526 
## itel    4 lambda   0.001000 stress 0.000001 penalty 0.204225 
## itel    2 lambda   0.002000 stress 0.000002 penalty 0.203535 
## itel    1 lambda   0.468000 stress 0.044604 penalty 0.000001 
## itel    1 lambda   0.469000 stress 0.044604 penalty 0.000000 
## itel  166 lambda   0.470000 stress 0.044603 penalty 0.000000
\end{verbatim}

\begin{center}\includegraphics{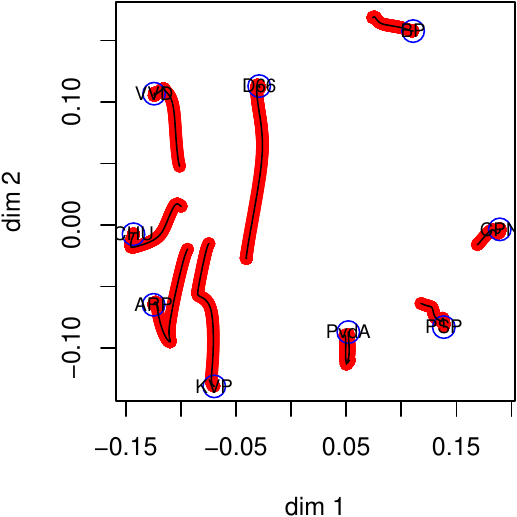} \end{center}

The singular values of the FDS solution are 2.95e-01, 2.10e-01,
1.89e-01, 1.34e-01, 1.16e-01, 1.06e-01, 8.61e-02, 7.06e-02, 1.10e-17,
and the Gower rank is probably eight. This is mainly because these data,
being averages, regress to the mean and thus have a substantial additive
constant. If we repeat the analysis after subtracting .1 from all
dissimilarities we get basically the same solution, but with somewhat
smoother trajectories.

\begin{verbatim}
## itel  511 lambda   0.000000 stress 0.000176 penalty 0.150789 
## itel    2 lambda   0.001000 stress 0.000176 penalty 0.037759 
## itel    2 lambda   0.002000 stress 0.000176 penalty 0.037619 
## itel    1 lambda   0.370000 stress 0.007642 penalty 0.000000 
## itel    1 lambda   0.371000 stress 0.007642 penalty 0.000000 
## itel    1 lambda   0.372000 stress 0.007642 penalty 0.000000
\end{verbatim}

\begin{center}\includegraphics{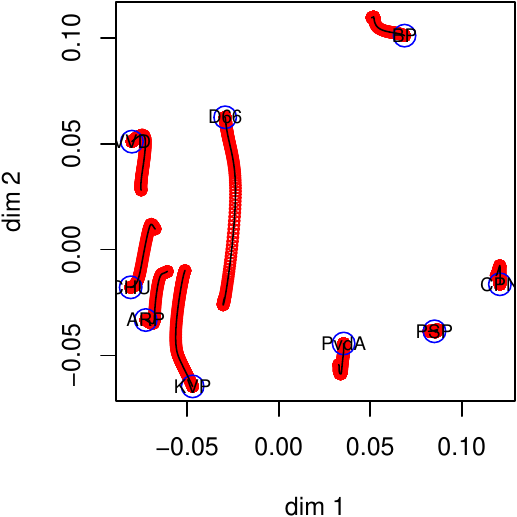} \end{center}

Now the singular values of the FDS solution are 2.05e-01, 1.34e-01,
1.11e-01, 6.03e-02, 3.11e-02, 3.97e-04, 1.18e-07, 4.55e-12, 1.16e-17,
and the approximate Gower rank is more like five or six.

\subsection{Ekman}\label{ekman}

The next example analyzes dissimilarities between 14 colors, taken from
Ekman (1954). The original similarities \(s_{ij}\), averaged over 31
subjects, were transformed to dissimilarities by
\(\delta_{ij}=1-s_{ij}\).

\begin{verbatim}
## itel 1482 lambda   0.000000 stress 0.000088 penalty 0.426110 
## itel    5 lambda   0.010000 stress 0.000132 penalty 0.118988 
## itel    3 lambda   0.020000 stress 0.000253 penalty 0.112777 
## itel    1 lambda   0.100000 stress 0.003195 penalty 0.070791 
## itel    1 lambda   0.200000 stress 0.010778 penalty 0.024407 
## itel    1 lambda   0.300000 stress 0.016125 penalty 0.003230 
## itel    1 lambda   0.400000 stress 0.017142 penalty 0.000165 
## itel    4 lambda   0.500000 stress 0.017213 penalty 0.000000 
## itel    1 lambda   0.600000 stress 0.017213 penalty 0.000000 
## itel    1 lambda   0.610000 stress 0.017213 penalty 0.000000 
## itel    1 lambda   0.620000 stress 0.017213 penalty 0.000000 
## itel    1 lambda   0.630000 stress 0.017213 penalty 0.000000
\end{verbatim}

\begin{center}\includegraphics{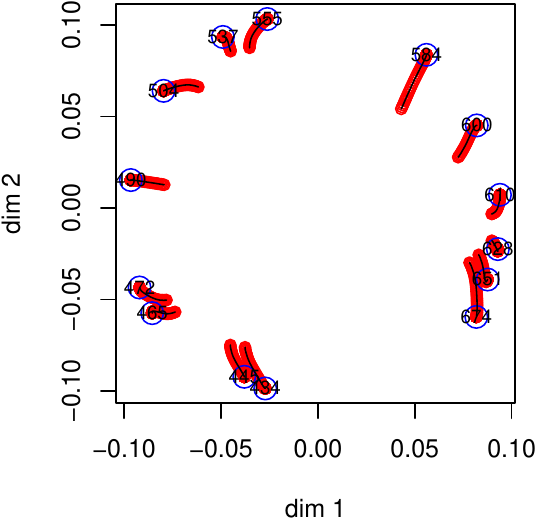} \end{center}

If we transform the Ekman similarities by \(\delta_{ij}=(1-s_{ij})^3\)
then its is known (De Leeuw (2016)) that the Gower rank is equal to two.
Thus the FDS solution has rank 2, and the 2MDS solution is the global
minimum.

\begin{verbatim}
## itel   99 lambda   0.000000 stress 0.011025 penalty 0.433456 
## itel    1 lambda   0.010000 stress 0.011025 penalty 0.000000 
## itel    1 lambda   0.020000 stress 0.011025 penalty 0.000000 
## itel    1 lambda   0.100000 stress 0.011025 penalty 0.000000 
## itel    1 lambda   0.110000 stress 0.011025 penalty 0.000000 
## itel    1 lambda   0.120000 stress 0.011025 penalty 0.000000 
## itel    1 lambda   0.130000 stress 0.011025 penalty 0.000000
\end{verbatim}

\begin{center}\includegraphics{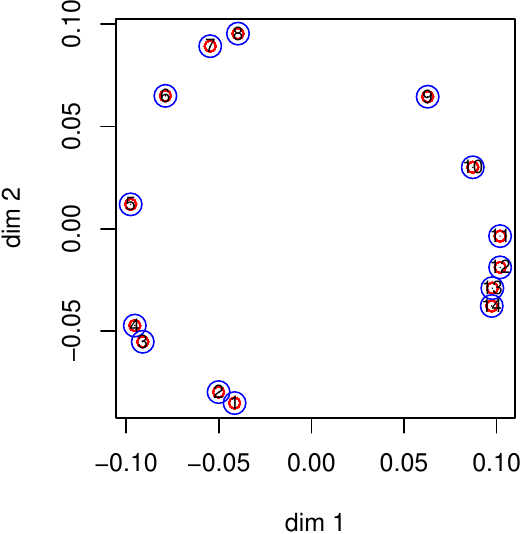} \end{center}

\subsection{Morse in Two}\label{morse-in-two}

Next, we use dissimilarities between 36 Morse code signals (Rothkopf
(1957)). We used the symmetrized version \texttt{morse} from the
\texttt{smacof} package (De Leeuw and Mair (2009)).

\begin{verbatim}
## itel 1461 lambda   0.000000 stress 0.000763 penalty 0.472254 
## itel    6 lambda   0.010000 stress 0.000858 penalty 0.322181 
## itel    4 lambda   0.020000 stress 0.001147 penalty 0.308335 
## itel    1 lambda   0.100000 stress 0.008576 penalty 0.216089 
## itel    1 lambda   0.200000 stress 0.028903 penalty 0.119364 
## itel    1 lambda   0.300000 stress 0.051285 penalty 0.060060 
## itel    1 lambda   0.400000 stress 0.068653 penalty 0.028190 
## itel    1 lambda   0.500000 stress 0.080258 penalty 0.011356 
## itel    1 lambda   0.600000 stress 0.086572 penalty 0.003578 
## itel    1 lambda   0.700000 stress 0.089140 penalty 0.000854 
## itel    1 lambda   0.800000 stress 0.089898 penalty 0.000116 
## itel    1 lambda   0.830000 stress 0.089958 penalty 0.000053 
## itel    1 lambda   0.840000 stress 0.089970 penalty 0.000040 
## itel  197 lambda   0.850000 stress 0.089949 penalty 0.000000
\end{verbatim}

\begin{center}\includegraphics{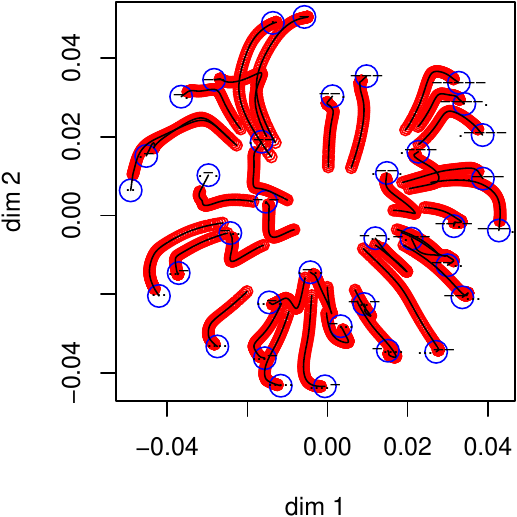} \end{center}

\subsection{Vegetables}\label{vegetables}

Our first one-dimensional example uses paired comparisons of 9
vegetables, originating with Guilford (1954) and taken from the
\texttt{psych} package (Revelle (2018)). The proportions are transformed
to dissimilarities by using the normal quantile function, i.e.
\(\delta_{ij}=|\Phi^{-1}(p_{ij})|\). We use a short sequence for
\(\lambda\).

\begin{verbatim}
## itel 1412 lambda   0.000000 stress 0.013675 penalty 0.269308 
## itel    5 lambda   0.010000 stress 0.013716 penalty 0.114786 
## itel    5 lambda   0.100000 stress 0.016719 penalty 0.069309 
## itel   23 lambda   1.000000 stress 0.035301 penalty 0.000000
\end{verbatim}

\begin{center}\includegraphics{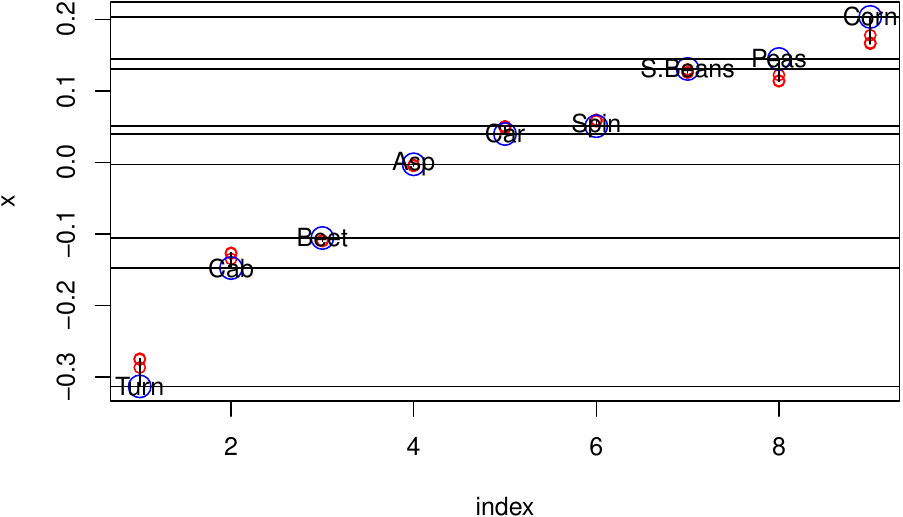} \end{center}

This example was previously analyzed in De Leeuw (2005) using
enumeration of all permutations. He found 14354 isolated local minima,
and a global minimum equal to the one we computed here.

\subsection{Plato}\label{plato}

Mair, Groenen, and De Leeuw (2022) use seriation of the works of Plato,
from the data collected by Cox and Brandwood (1959), as an example of
unidimensional scaling. We first run this example with our usual
sequence of five \(\lambda\) values.

\begin{verbatim}
## itel  169 lambda   0.000000 stress 0.000000 penalty 0.410927 
## itel    3 lambda   0.010000 stress 0.000062 penalty 0.255246 
## itel    3 lambda   0.100000 stress 0.005117 penalty 0.194993 
## itel    4 lambda   1.000000 stress 0.106058 penalty 0.019675 
## itel    9 lambda  10.000000 stress 0.139462 penalty 0.000000
\end{verbatim}

\begin{center}\includegraphics{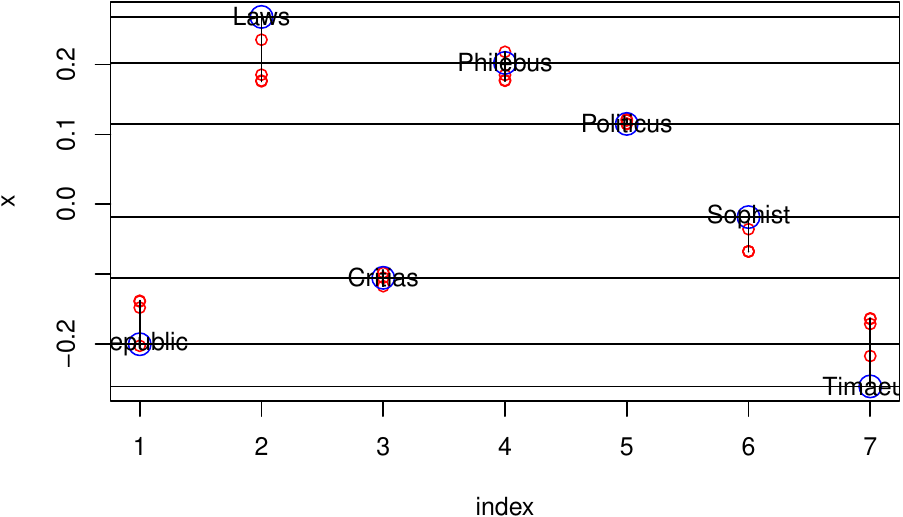} \end{center}

This gives the order

\begin{verbatim}
##      [,1]       
## [1,] "Timaeus"  
## [2,] "Republic" 
## [3,] "Critias"  
## [4,] "Sophist"  
## [5,] "Politicus"
## [6,] "Philebus" 
## [7,] "Laws"
\end{verbatim}

which is different from the order at the global minimum, which has
Republic before Timaeus. Thus we have recovered a local minimum, and it
seems our sequence of \(\lambda\) values was not fine enough to do the
job properly. So we try a longer and finer sequence.

\begin{verbatim}
## itel  169 lambda   0.000000 stress 0.000000 penalty 0.410927 
## itel    3 lambda   0.000100 stress 0.000000 penalty 0.263015 
## itel    3 lambda   0.001000 stress 0.000001 penalty 0.262280 
## itel    3 lambda   0.010000 stress 0.000064 penalty 0.255078 
## itel    3 lambda   0.100000 stress 0.005123 penalty 0.194945 
## itel    2 lambda   0.200000 stress 0.016184 penalty 0.147493 
## itel    1 lambda   0.300000 stress 0.026997 penalty 0.119323 
## itel    1 lambda   0.400000 stress 0.040023 penalty 0.093615 
## itel    1 lambda   0.500000 stress 0.053688 penalty 0.072330 
## itel    1 lambda   0.600000 stress 0.066833 penalty 0.055452 
## itel    1 lambda   0.700000 stress 0.078832 penalty 0.042269 
## itel    1 lambda   0.800000 stress 0.089439 penalty 0.032019 
## itel    1 lambda   0.900000 stress 0.098557 penalty 0.024079 
## itel    1 lambda   1.000000 stress 0.106135 penalty 0.017940 
## itel    6 lambda   2.000000 stress 0.130789 penalty 0.000148 
## itel   13 lambda   3.000000 stress 0.131135 penalty 0.000000
\end{verbatim}

\begin{center}\includegraphics{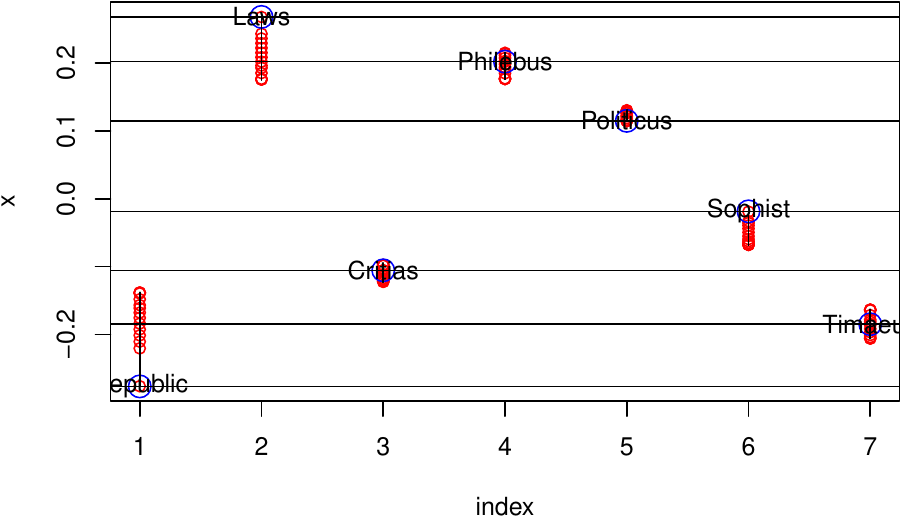} \end{center}

Now the order is

\begin{verbatim}
##      [,1]       
## [1,] "Republic" 
## [2,] "Timaeus"  
## [3,] "Critias"  
## [4,] "Sophist"  
## [5,] "Politicus"
## [6,] "Philebus" 
## [7,] "Laws"
\end{verbatim}

which does indeed correspond to the global minimum.

With a different \(\lambda\) sequence we find the same solution.

\begin{verbatim}
## itel  169 lambda   0.000000 stress 0.000000 penalty 0.410927 
## itel    3 lambda   0.001000 stress 0.000001 penalty 0.262296 
## itel    2 lambda   0.002000 stress 0.000003 penalty 0.261483 
## itel    2 lambda   0.004000 stress 0.000010 penalty 0.259872 
## itel    2 lambda   0.008000 stress 0.000041 penalty 0.256690 
## itel    2 lambda   0.016000 stress 0.000159 penalty 0.250470 
## itel    2 lambda   0.032000 stress 0.000613 penalty 0.238574 
## itel    2 lambda   0.064000 stress 0.002266 penalty 0.216785 
## itel    2 lambda   0.128000 stress 0.007791 penalty 0.180067 
## itel    2 lambda   0.256000 stress 0.023483 penalty 0.127006 
## itel    2 lambda   0.512000 stress 0.056940 penalty 0.067948 
## itel    3 lambda   1.024000 stress 0.107743 penalty 0.017937 
## itel    8 lambda   2.048000 stress 0.131059 penalty 0.000032 
## itel    9 lambda   4.096000 stress 0.131135 penalty 0.000000
\end{verbatim}

\begin{center}\includegraphics{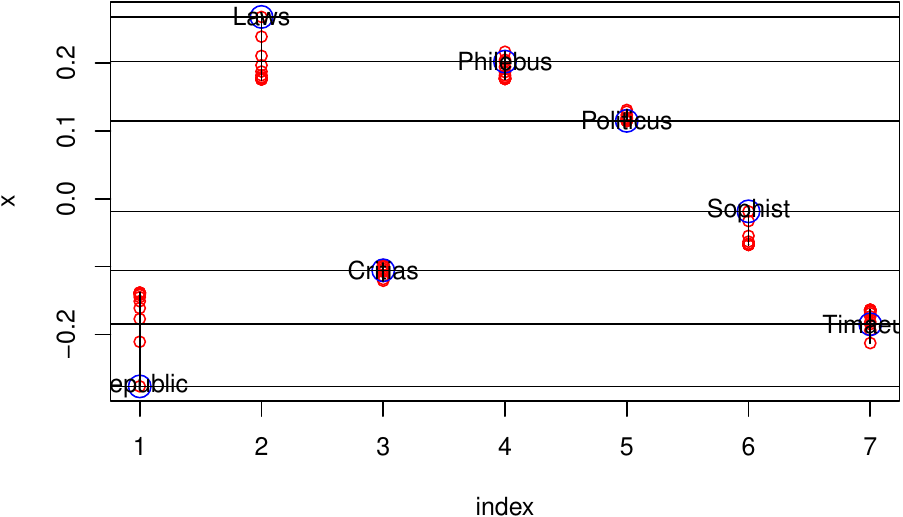} \end{center}

The order is

\begin{verbatim}
##      [,1]       
## [1,] "Republic" 
## [2,] "Timaeus"  
## [3,] "Critias"  
## [4,] "Sophist"  
## [5,] "Politicus"
## [6,] "Philebus" 
## [7,] "Laws"
\end{verbatim}

\subsection{Morse in One}\label{morse-in-one}

Now for a more challenging example. The Morse code data have been used
to try out exact unidimensional MDS techniques, for example by
Palubeckis (2013). We will enter the global minimum contest by using
10,000 values of \(\lambda\), in an equally spaced sequence from 0 to
10. This is not as bad as it sounds. For the 10,000 FDS solutions
\texttt{system.time()} tells us

\begin{verbatim}
##    user  system elapsed 
##   4.688   0.216   4.906
\end{verbatim}

The one-dimensional plot show quite a bit of movement, but much of it
seems to be contained in the very first change of \(\lambda\).

\begin{center}\includegraphics[width=0.85\linewidth]{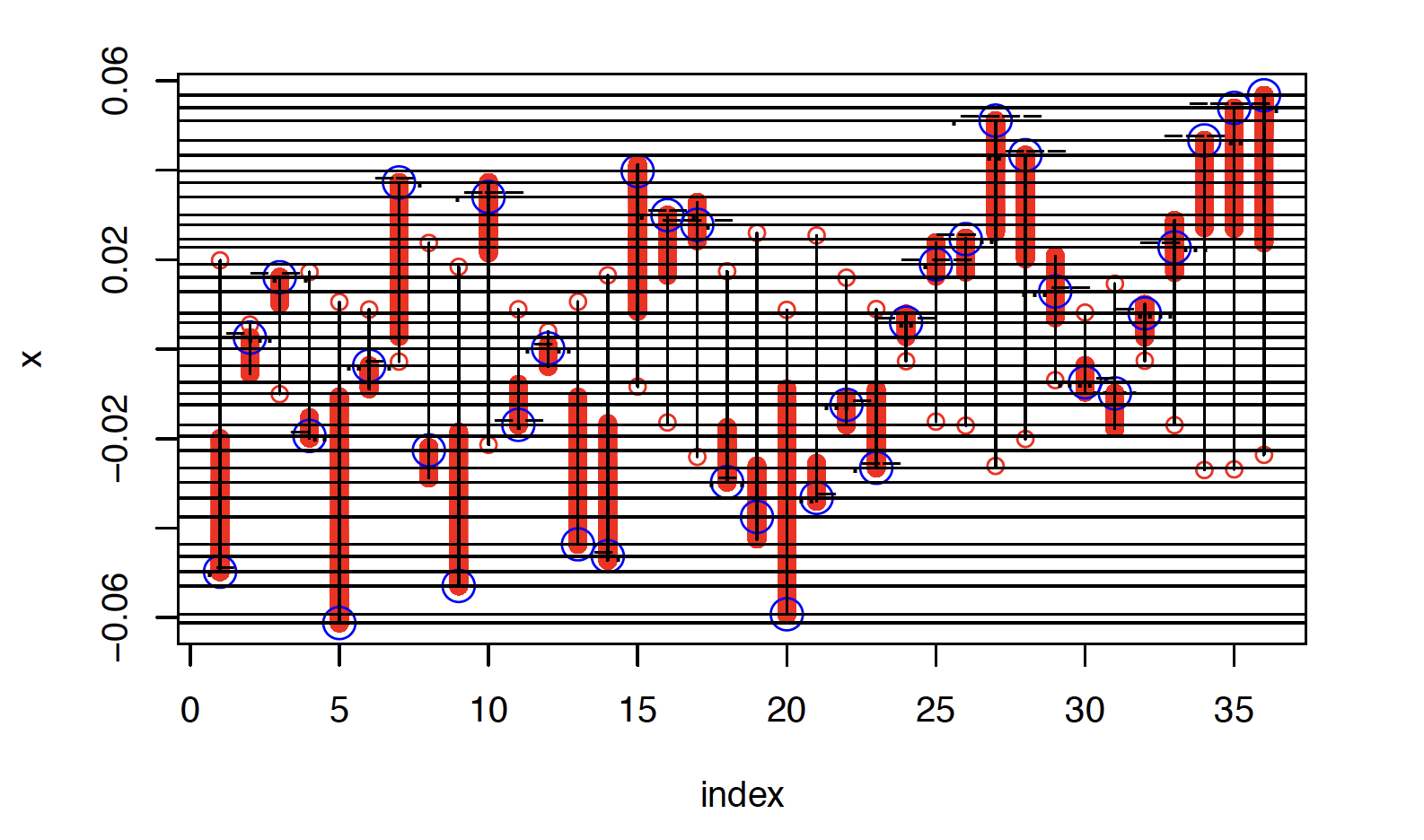} \end{center}

We can also plot stress and the penalty term as functions of
\(\lambda\). Again, note the big change in the penalty term when
\(\lambda\) goes from zero to 0.001.

\begin{center}\includegraphics{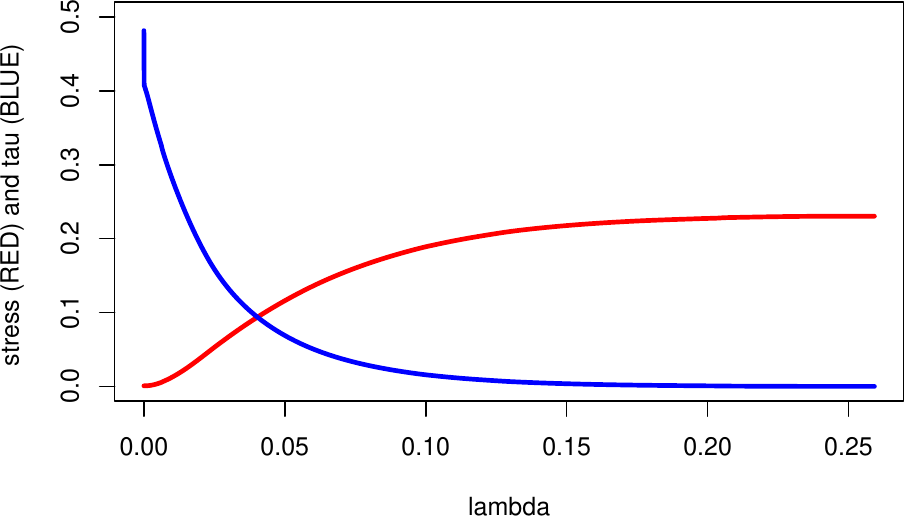} \end{center}

After the first 2593 values of \(\lambda\) the penalty term is zero and
we stop, i.e.~we estimate \(\lambda_+\) is 2.593. At that point we have
run a total of 5013 FDS iterations, and thus on average about two
iterations per \(\lambda\) value. Stress has increased from 0.0007634501
to 0.2303106976 and the penalty value has decreased from 0.4815136419 to
0.0000000001. We find the following order of the points on the
dimension.

\begin{verbatim}
##       [,1]   
##  [1,] "."    
##  [2,] "-"    
##  [3,] ".."   
##  [4,] ".-"   
##  [5,] "-."   
##  [6,] "--"   
##  [7,] "..."  
##  [8,] "..-"  
##  [9,] ".-."  
## [10,] ".--"  
## [11,] "...." 
## [12,] "-.."  
## [13,] "-.-"  
## [14,] "...-" 
## [15,] "....."
## [16,] "....-"
## [17,] "..-." 
## [18,] ".-.." 
## [19,] "-..." 
## [20,] "-..-" 
## [21,] "-...."
## [22,] "...--"
## [23,] "-.-." 
## [24,] "-.--" 
## [25,] "--..."
## [26,] "--.." 
## [27,] "--.-" 
## [28,] ".--." 
## [29,] ".---" 
## [30,] "--."  
## [31,] "---"  
## [32,] "..---"
## [33,] "---.."
## [34,] ".----"
## [35,] "----."
## [36,] "-----"
\end{verbatim}

Our order, and consequently our solution, is the same as the exact
global solution given by Palubeckis (2013). See his table 4, reproduced
below. The difference is that computing our solution takes 10 seconds,
while his takes 494 seconds. But of course we would not know we actually
found the global mimimum if the exact exhaustive methods had not
analyzed the same data before.

\begin{center}\includegraphics[width=0.75\linewidth]{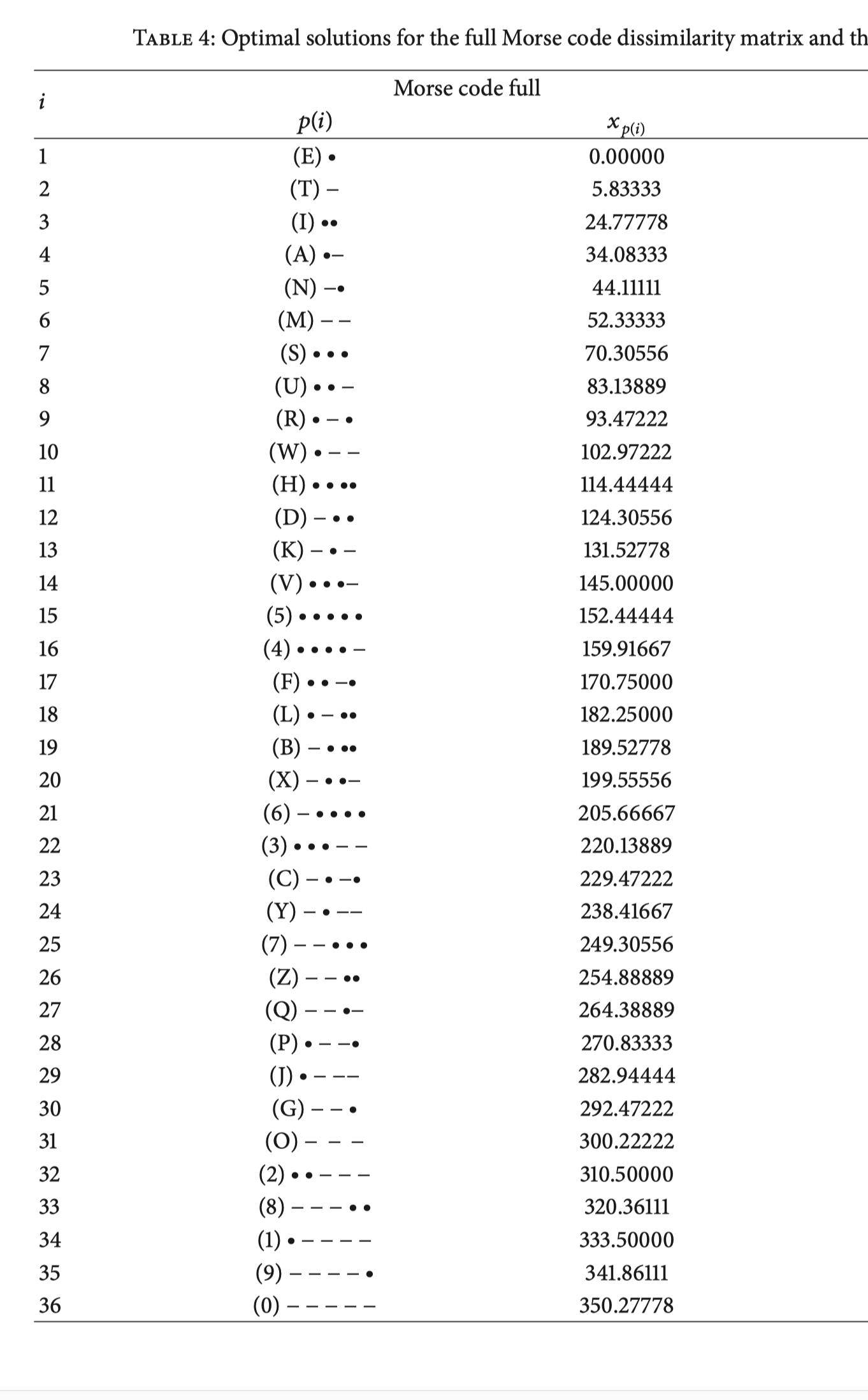} \end{center}

\section{Discussion}\label{discussion}

There is one surprising (to me, at least) finding from all our examples.
There is a value, say \(\lambda_+\), such that the penalty \(\tau(Y)\)
is zero for all PFDS solutions with \(\lambda\geq\lambda_+\). In other
words, our penalty function acts like a \emph{smooth exact penalty
function}. The precise reason for exactness in our case (if there is
one) is not entirely clear to me yet, but it is obviously a topic for
further research, using for example the recent theoretical framework of
Dolgopolik (2016a), Dolgopolik (2016b), Dolgopolik (2017), Dolgopolik
(n.d.).

In our two dimensional examples we always start our plots with the first
two dimensions of the FDS configuration. These two-dimensional
configurations are usually small (all points relatively close to the
origin), because so much variation is still in the higher dimensions. If
\(\lambda\) increases the growth of the configurations is one important
aspect of configuration change.

In our iteration counts with short sequences of \(\lambda\) we see
relatively small increases in stress and small decreases in the penalty
term, until we get closer to \(\lambda_+\), when we suddenly see a
sudden change and a larger number of iterations. This is also reflected
in the figures, where generally the change to the last solution (with
the largest \(\lambda\)) makes the largest jump. This suggest a finer
sequence near \(\lambda_+\) and perhaps an adaptive strategy for
choosing \(\lambda\). Or to use brute force, as in the unidimensional
Morse code example. With such longer and finer sequences convergence
becomes more smooth.

Another all-important aspect of the method discussed here is that it
assumes computation of the global minimum for each \(\lambda\). Since we
cannot expect a result as nice as the one for FDS (all local minima are
global) for \(\lambda>0\) our method remains somewhat heuristic. We have
seen that some sequences of \(\lambda\) can take us to a non-global
local minimum. Of course the fact that we start with a global minimum
for \(\lambda=0\) is of some help, but we do not know how far it will
take us in general. Jumps near \(\lambda_+\) may indicate bifurcations
to other local minima.

We have not stressed in the paper that minimizing the penalty function
is a \emph{continuation method} (Allgower and George (1979)). This means
that probably better methods are available to follow the trajectory of
solutions along \(\lambda>0\). There are also possibilities in exploring
the fact that the maximum over \(Z\) of the penalty function
\(\eqref{E:pi}\) is a concave function of the single variable
\(\lambda\), which is a constant function for all \(\lambda>\lambda_+\).
There is a duality theory associated with these Courant penalty
functions, which we have not used or explored so far.

\section{Appendix A: Exterior Penalty
Methods}\label{appendix-a-exterior-penalty-methods}

Suppose \(\mathcal{X}\subseteq\mathbb{R}^n\) and
\(f:\mathbb{R}^n\Rightarrow\mathbb{R}\) is continuous. Define \[
\mathcal{X}_\star=\mathop{\text{argmin}}_{x\in\mathcal{X}}\ f(x)
\] Suppose \(\mathcal{X}_\star\) is non-empty and that \(x_\star\) is
any element of \(\mathcal{X}_\star\), and \[
f_\star=f(x_\star)=\min_{x\in\mathcal{X}}\ f(x).
\]

The following convergence analysis of external linear penalty methods is
standard and can be found in many texts (for example, Zangwill (1969),
section 12.2).

The penalty term \(g:\mathbb{R}^n\Rightarrow\mathbb{R}^+\) is continuous
and satisfies \(g(x)=0\) if and only if \(x\in\mathcal{X}\). For each
\(\lambda>0\) we define the (linear, external) penalty function
\begin{equation}
h(x,\lambda)=f(x)+\lambda g(x).
\end{equation}

Suppose \(\{\lambda_k\}\) is a strictly increasing sequence of positive
real numbers. Define \begin{equation}
\mathcal{X}_k=\mathop{\text{argmin}}_{x\in\mathcal{X}}\ h(x,\lambda_k).
\end{equation} Suppose all \(\mathcal{X}_k\) are nonempty and contained
in a compact subset of \(\mathcal{X}\). Choose \(x_k\in\mathcal{X}_k\)
arbitrarily.

\textbf{Lemma 2: {[}Basic{]}}

1: \(h(x_k,\lambda_k)\leq h(x_{k+1},\lambda_{k+1})\).

2: \(g(x_k)\geq g(x_{k+1})\).

3: \(f(x_k)\leq f(x_{k+1})\).

4: \(f_\star\geq h(x_k,\lambda_k)\geq f(x_k)\).

\textbf{Proof:}

1: We have the chain \[
h(x_{k+1},\lambda_{k+1})=f(x_{k+1})+\lambda_{k+1} g(x_{k+1})\geq f(x_{k+1})+\lambda_{k} g(x_{k+1})\geq f(x_{k})+\lambda_{k}g(x_k)=h(x_k,\lambda_k).
\] 2: Both \begin{align}
f(x_k)+\lambda_k g(x_k)&\leq f(x_{k+1})+\lambda_k g(x_{k+1}),\label{E:21}\\
f(x_{k+1})+\lambda_{k+1} g(x_{k+1})&\leq f(x_k)+\lambda_{k+1} g(x_k).\label{E:22}
\end{align} Adding inequalities \(\eqref{E:21}\) and \(\eqref{E:22}\)
gives \[
\lambda_k g(x_k)+\lambda_{k+1} g(x_{k+1})\leq\lambda_k g(x_{k+1})+\lambda_{k+1} g(x_k),
\] or \[
(\lambda_k-\lambda_{k+1})g(x_k)\leq(\lambda_k-\lambda_{k+1})g(x_{k+1}),
\] and thus \(g(x_k)\geq g(x_{k+1})\).

3: First \begin{equation}\label{E:31}
f(x_{k+1})+\lambda_k g(x_{k+1})\geq f(x_k)+\lambda_k g(x_k).
\end{equation} We just proved that \(g(x_{k+1})\geq g(x_k)\), and thus
\begin{equation}\label{E:32}
f(x_k)+\lambda_k g(x_k)\geq f(x_k)+\lambda_k g(x_{k+1}).
\end{equation} Combining inequalities \(\eqref{E:31}\) and
\(\eqref{E:32}\) gives \(f(x_{k+1})\geq f(x_k)\).

4: We have the chain \[
f_\star=f(x_\star)+\lambda_k g(x_\star)\geq f(x_k)+\lambda_k g(x_k)\geq f(x_k).
\] \(\blacksquare\)

\textbf{Theorem 3: } Suppose the sequence \(\{\lambda_k\}_{k\in K}\)
diverges to \(\infty\) and \(x_{\star\star}\) is the limit of any
convergent subsequence \(\{x_\ell\}_{\ell\in L}\). Then
\(x_{\star\star}\in\mathcal{X}_\star\), and
\(f(x_{\star\star})=f_\star\), and \(g(x_{\star\star})=0\).

\textbf{Proof:} Using part 4 of lemma 2 \[
\lim_{\ell\in L}h(x_\ell,\lambda_\ell)=\lim_{\ell\in L}\{f(x_\ell)+\lambda_\ell g(x_\ell)\}=f(x_{\star\star})+\lim_{\ell\in L}\lambda_\ell g(x_\ell)\leq f(x_\star).
\]

Thus \(\{h(x_\ell,\lambda_\ell)_{\ell\in L}\}\) is a bounded increasing
sequence, which consequently converges, and
\(\lim_{\ell\in L}\lambda_\ell g(x_\ell)\) also converges. Since
\(\{\lambda_\ell\}_{\ell\in L}\rightarrow\infty\) it follows that
\(\lim_{\ell\in L}g(x_\ell)=g(x_{\star\star})=0\). Thus
\(x_{\star\star}\in\mathcal{X}\). Since \(f(x_\ell)\leq f_\star\) we see
that \(f(x_{\star\star})\leq f_\star\), and thus
\(x_{\star\star}\in\mathcal{X}_\star\) and
\(f(x_{\star\star})=f_\star\). \(\blacksquare\)

\section{Appendix B: Code}\label{appendix-b-code}

\subsection{penalty.R}\label{penalty.r}

\begin{Shaded}
\begin{Highlighting}[]
\NormalTok{smacofLoss }\OtherTok{\textless{}{-}} \ControlFlowTok{function}\NormalTok{ (d, w, delta) \{}
  \FunctionTok{return}\NormalTok{ (}\FunctionTok{sum}\NormalTok{ (w }\SpecialCharTok{*}\NormalTok{ (delta }\SpecialCharTok{{-}}\NormalTok{ d) }\SpecialCharTok{\^{}} \DecValTok{2}\NormalTok{) }\SpecialCharTok{/} \DecValTok{4}\NormalTok{)}
\NormalTok{\}}

\NormalTok{smacofBmat }\OtherTok{\textless{}{-}} \ControlFlowTok{function}\NormalTok{ (d, w, delta) \{}
\NormalTok{  dd }\OtherTok{\textless{}{-}} \FunctionTok{ifelse}\NormalTok{ (d }\SpecialCharTok{==} \DecValTok{0}\NormalTok{, }\DecValTok{0}\NormalTok{, }\DecValTok{1} \SpecialCharTok{/}\NormalTok{ d)}
\NormalTok{  b }\OtherTok{\textless{}{-}} \SpecialCharTok{{-}}\NormalTok{dd }\SpecialCharTok{*}\NormalTok{ w }\SpecialCharTok{*}\NormalTok{ delta}
  \FunctionTok{diag}\NormalTok{ (b) }\OtherTok{\textless{}{-}} \SpecialCharTok{{-}}\FunctionTok{rowSums}\NormalTok{ (b)}
  \FunctionTok{return}\NormalTok{(b)}
\NormalTok{\}}

\NormalTok{smacofVmat }\OtherTok{\textless{}{-}} \ControlFlowTok{function}\NormalTok{ (w) \{}
\NormalTok{  v }\OtherTok{\textless{}{-}} \SpecialCharTok{{-}}\NormalTok{w}
  \FunctionTok{diag}\NormalTok{(v) }\OtherTok{\textless{}{-}} \SpecialCharTok{{-}}\FunctionTok{rowSums}\NormalTok{(v)}
  \FunctionTok{return}\NormalTok{ (v)}
\NormalTok{\}}

\NormalTok{smacofGuttman }\OtherTok{\textless{}{-}} \ControlFlowTok{function}\NormalTok{ (x, b, vinv) \{}
  \FunctionTok{return}\NormalTok{ (vinv }\SpecialCharTok{\%*\%}\NormalTok{ b }\SpecialCharTok{\%*\%}\NormalTok{ x)}
\NormalTok{\}}

\NormalTok{columnCenter }\OtherTok{\textless{}{-}} \ControlFlowTok{function}\NormalTok{ (x) \{}
  \FunctionTok{return}\NormalTok{ (}\FunctionTok{apply}\NormalTok{ (x, }\DecValTok{2}\NormalTok{, }\ControlFlowTok{function}\NormalTok{ (z) z }\SpecialCharTok{{-}} \FunctionTok{mean}\NormalTok{ (z)))}
\NormalTok{\}}

\NormalTok{smacofComplement }\OtherTok{\textless{}{-}} \ControlFlowTok{function}\NormalTok{ (y, v) \{}
  \FunctionTok{return}\NormalTok{ (}\FunctionTok{sum}\NormalTok{ (v }\SpecialCharTok{*} \FunctionTok{tcrossprod}\NormalTok{ (y)) }\SpecialCharTok{/} \DecValTok{4}\NormalTok{)}
\NormalTok{\}}

\NormalTok{smacofPenalty }\OtherTok{\textless{}{-}}
  \ControlFlowTok{function}\NormalTok{ (w,}
\NormalTok{            delta,}
            \AttributeTok{p =} \DecValTok{2}\NormalTok{,}
            \AttributeTok{lbd =} \DecValTok{0}\NormalTok{,}
            \AttributeTok{zold =} \FunctionTok{columnCenter}\NormalTok{ (}\FunctionTok{diag}\NormalTok{ (}\FunctionTok{nrow}\NormalTok{ (delta))),}
            \AttributeTok{itmax =} \DecValTok{10000}\NormalTok{,}
            \AttributeTok{eps =} \FloatTok{1e{-}10}\NormalTok{,}
            \AttributeTok{verbose =} \ConstantTok{FALSE}\NormalTok{) \{}
\NormalTok{    itel }\OtherTok{\textless{}{-}} \DecValTok{1}
\NormalTok{    n }\OtherTok{\textless{}{-}} \FunctionTok{nrow}\NormalTok{ (zold)}
\NormalTok{    vmat }\OtherTok{\textless{}{-}} \FunctionTok{smacofVmat}\NormalTok{ (w)}
\NormalTok{    vinv }\OtherTok{\textless{}{-}} \FunctionTok{solve}\NormalTok{ (vmat }\SpecialCharTok{+}\NormalTok{ (}\DecValTok{1} \SpecialCharTok{/}\NormalTok{ n)) }\SpecialCharTok{{-}}\NormalTok{ (}\DecValTok{1} \SpecialCharTok{/}\NormalTok{ n)}
\NormalTok{    dold }\OtherTok{\textless{}{-}} \FunctionTok{as.matrix}\NormalTok{ (}\FunctionTok{dist}\NormalTok{ (zold))}
\NormalTok{    mold }\OtherTok{\textless{}{-}} \FunctionTok{sum}\NormalTok{ (w }\SpecialCharTok{*}\NormalTok{ delta }\SpecialCharTok{*}\NormalTok{ dold) }\SpecialCharTok{/} \FunctionTok{sum}\NormalTok{ (w }\SpecialCharTok{*}\NormalTok{ dold }\SpecialCharTok{*}\NormalTok{ dold)}
\NormalTok{    zold }\OtherTok{\textless{}{-}}\NormalTok{ zold }\SpecialCharTok{*}\NormalTok{ mold}
\NormalTok{    dold }\OtherTok{\textless{}{-}}\NormalTok{ dold }\SpecialCharTok{*}\NormalTok{ mold}
\NormalTok{    yold }\OtherTok{\textless{}{-}}\NormalTok{ zold [, (p }\SpecialCharTok{+} \DecValTok{1}\NormalTok{) }\SpecialCharTok{:}\NormalTok{ n]}
\NormalTok{    sold }\OtherTok{\textless{}{-}} \FunctionTok{smacofLoss}\NormalTok{ (dold, w, delta)}
\NormalTok{    bold }\OtherTok{\textless{}{-}} \FunctionTok{smacofBmat}\NormalTok{ (dold, w, delta)}
\NormalTok{    told }\OtherTok{\textless{}{-}} \FunctionTok{smacofComplement}\NormalTok{ (yold, vmat)}
\NormalTok{    uold }\OtherTok{\textless{}{-}}\NormalTok{ sold }\SpecialCharTok{+}\NormalTok{ lbd }\SpecialCharTok{*}\NormalTok{ told}
    \ControlFlowTok{repeat}\NormalTok{ \{}
\NormalTok{      znew }\OtherTok{\textless{}{-}} \FunctionTok{smacofGuttman}\NormalTok{ (zold, bold, vinv)}
\NormalTok{      ynew }\OtherTok{\textless{}{-}}\NormalTok{ znew [, (p }\SpecialCharTok{+} \DecValTok{1}\NormalTok{) }\SpecialCharTok{:}\NormalTok{ n] }\SpecialCharTok{/}\NormalTok{ (}\DecValTok{1} \SpecialCharTok{+}\NormalTok{ lbd)}
\NormalTok{      znew [, (p }\SpecialCharTok{+} \DecValTok{1}\NormalTok{) }\SpecialCharTok{:}\NormalTok{ n] }\OtherTok{\textless{}{-}}\NormalTok{ ynew}
\NormalTok{      xnew }\OtherTok{\textless{}{-}}\NormalTok{ znew [, }\DecValTok{1}\SpecialCharTok{:}\NormalTok{p]}
\NormalTok{      dnew }\OtherTok{\textless{}{-}} \FunctionTok{as.matrix}\NormalTok{ (}\FunctionTok{dist}\NormalTok{ (znew))}
\NormalTok{      bnew }\OtherTok{\textless{}{-}} \FunctionTok{smacofBmat}\NormalTok{ (dnew, w, delta)}
\NormalTok{      tnew }\OtherTok{\textless{}{-}} \FunctionTok{smacofComplement}\NormalTok{ (ynew, vmat)}
\NormalTok{      snew }\OtherTok{\textless{}{-}} \FunctionTok{smacofLoss}\NormalTok{ (dnew, w, delta) }
\NormalTok{      unew }\OtherTok{\textless{}{-}}\NormalTok{ snew }\SpecialCharTok{+}\NormalTok{ lbd }\SpecialCharTok{*}\NormalTok{ tnew }
      \ControlFlowTok{if}\NormalTok{ (verbose) \{}
        \FunctionTok{cat}\NormalTok{(}
          \StringTok{"itel "}\NormalTok{,}
          \FunctionTok{formatC}\NormalTok{(itel, }\AttributeTok{width =} \DecValTok{4}\NormalTok{, }\AttributeTok{format =} \StringTok{"d"}\NormalTok{),}
          \StringTok{"sold "}\NormalTok{,}
          \FunctionTok{formatC}\NormalTok{(}
\NormalTok{            sold,}
            \AttributeTok{width =} \DecValTok{10}\NormalTok{,}
            \AttributeTok{digits =} \DecValTok{6}\NormalTok{,}
            \AttributeTok{format =} \StringTok{"f"}
\NormalTok{          ),}
          \StringTok{"snew "}\NormalTok{,}
          \FunctionTok{formatC}\NormalTok{(}
\NormalTok{            snew,}
            \AttributeTok{width =} \DecValTok{10}\NormalTok{,}
            \AttributeTok{digits =} \DecValTok{6}\NormalTok{,}
            \AttributeTok{format =} \StringTok{"f"}
\NormalTok{          ),}
          \StringTok{"told "}\NormalTok{,}
          \FunctionTok{formatC}\NormalTok{(}
\NormalTok{            told,}
            \AttributeTok{width =} \DecValTok{10}\NormalTok{,}
            \AttributeTok{digits =} \DecValTok{6}\NormalTok{,}
            \AttributeTok{format =} \StringTok{"f"}
\NormalTok{          ),}
          \StringTok{"tnew "}\NormalTok{,}
          \FunctionTok{formatC}\NormalTok{(}
\NormalTok{            tnew,}
            \AttributeTok{width =} \DecValTok{10}\NormalTok{,}
            \AttributeTok{digits =} \DecValTok{6}\NormalTok{,}
            \AttributeTok{format =} \StringTok{"f"}
\NormalTok{          ),}
          \StringTok{"uold "}\NormalTok{,}
          \FunctionTok{formatC}\NormalTok{(}
\NormalTok{            uold,}
            \AttributeTok{width =} \DecValTok{10}\NormalTok{,}
            \AttributeTok{digits =} \DecValTok{6}\NormalTok{,}
            \AttributeTok{format =} \StringTok{"f"}
\NormalTok{          ),}
          \StringTok{"unew "}\NormalTok{,}
          \FunctionTok{formatC}\NormalTok{(}
\NormalTok{            unew,}
            \AttributeTok{width =} \DecValTok{10}\NormalTok{,}
            \AttributeTok{digits =} \DecValTok{6}\NormalTok{,}
            \AttributeTok{format =} \StringTok{"f"}
\NormalTok{          ),}
          \StringTok{"}\SpecialCharTok{\textbackslash{}n}\StringTok{"}
\NormalTok{        )}
\NormalTok{      \}}
      \ControlFlowTok{if}\NormalTok{ (((uold }\SpecialCharTok{{-}}\NormalTok{ unew) }\SpecialCharTok{\textless{}}\NormalTok{ eps) }\SpecialCharTok{||}\NormalTok{ (itel }\SpecialCharTok{==}\NormalTok{ itmax)) \{}
        \ControlFlowTok{break}
\NormalTok{      \}}
\NormalTok{      itel }\OtherTok{\textless{}{-}}\NormalTok{ itel }\SpecialCharTok{+} \DecValTok{1}
\NormalTok{      zold }\OtherTok{\textless{}{-}}\NormalTok{ znew}
\NormalTok{      bold }\OtherTok{\textless{}{-}}\NormalTok{ bnew}
\NormalTok{      sold }\OtherTok{\textless{}{-}}\NormalTok{ snew}
\NormalTok{      told }\OtherTok{\textless{}{-}}\NormalTok{ tnew}
\NormalTok{      uold }\OtherTok{\textless{}{-}}\NormalTok{ unew}
\NormalTok{    \}}
\NormalTok{    zpri }\OtherTok{\textless{}{-}}\NormalTok{znew }\SpecialCharTok{\%*\%} \FunctionTok{svd}\NormalTok{(znew)}\SpecialCharTok{$}\NormalTok{v}
\NormalTok{    xpri }\OtherTok{\textless{}{-}}\NormalTok{ zpri[, }\DecValTok{1}\SpecialCharTok{:}\NormalTok{p]}
    \FunctionTok{return}\NormalTok{ (}\FunctionTok{list}\NormalTok{ (}
      \AttributeTok{x =}\NormalTok{ xpri,}
      \AttributeTok{z =}\NormalTok{ zpri,}
      \AttributeTok{b =}\NormalTok{ bnew,}
      \AttributeTok{l =}\NormalTok{ lbd,}
      \AttributeTok{s =}\NormalTok{ snew,}
      \AttributeTok{t =}\NormalTok{ tnew,}
      \AttributeTok{itel =}\NormalTok{ itel}
\NormalTok{    ))}
\NormalTok{  \}}
\end{Highlighting}
\end{Shaded}

\subsection{runPenalty.R}\label{runpenalty.r}

\begin{Shaded}
\begin{Highlighting}[]
\NormalTok{runPenalty }\OtherTok{\textless{}{-}}
  \ControlFlowTok{function}\NormalTok{ (w,}
\NormalTok{            delta,}
\NormalTok{            lbd,}
            \AttributeTok{p =} \DecValTok{2}\NormalTok{,}
            \AttributeTok{itmax =} \DecValTok{10000}\NormalTok{,}
            \AttributeTok{eps =} \FloatTok{1e{-}10}\NormalTok{,}
            \AttributeTok{cut =} \FloatTok{1e{-}6}\NormalTok{,}
            \AttributeTok{write =} \ConstantTok{TRUE}\NormalTok{,}
            \AttributeTok{verbose =} \ConstantTok{FALSE}\NormalTok{) \{}
\NormalTok{    m }\OtherTok{\textless{}{-}} \FunctionTok{length}\NormalTok{ (lbd)}
\NormalTok{    hList }\OtherTok{\textless{}{-}} \FunctionTok{as.list}\NormalTok{ (}\DecValTok{1}\SpecialCharTok{:}\NormalTok{m)}
\NormalTok{    hList[[}\DecValTok{1}\NormalTok{]] }\OtherTok{\textless{}{-}}
      \FunctionTok{smacofPenalty}\NormalTok{(}
\NormalTok{        w,}
\NormalTok{        delta,}
\NormalTok{        p,}
        \AttributeTok{lbd =}\NormalTok{ lbd[}\DecValTok{1}\NormalTok{],}
        \AttributeTok{itmax =}\NormalTok{ itmax,}
        \AttributeTok{eps =}\NormalTok{ eps,}
        \AttributeTok{verbose =}\NormalTok{ verbose}
\NormalTok{      )}
    \ControlFlowTok{for}\NormalTok{ (j }\ControlFlowTok{in} \DecValTok{2}\SpecialCharTok{:}\NormalTok{m) \{}
\NormalTok{      hList[[j]] }\OtherTok{\textless{}{-}}
        \FunctionTok{smacofPenalty}\NormalTok{(}
\NormalTok{          w,}
\NormalTok{          delta,}
\NormalTok{          p,}
          \AttributeTok{zold =}\NormalTok{ hList[[j }\SpecialCharTok{{-}} \DecValTok{1}\NormalTok{]]}\SpecialCharTok{$}\NormalTok{z,}
          \AttributeTok{lbd =}\NormalTok{ lbd[j],}
          \AttributeTok{itmax =}\NormalTok{ itmax,}
          \AttributeTok{eps =}\NormalTok{ eps,}
          \AttributeTok{verbose =}\NormalTok{ verbose}
\NormalTok{        )}
\NormalTok{    \}}
\NormalTok{    mm }\OtherTok{\textless{}{-}}\NormalTok{ m}
    \ControlFlowTok{for}\NormalTok{ (i }\ControlFlowTok{in} \DecValTok{1}\SpecialCharTok{:}\NormalTok{m) \{}
      \ControlFlowTok{if}\NormalTok{ (write) \{}
        \FunctionTok{cat}\NormalTok{(}
          \StringTok{"itel"}\NormalTok{,}
          \FunctionTok{formatC}\NormalTok{(hList[[i]]}\SpecialCharTok{$}\NormalTok{itel, }\AttributeTok{width =} \DecValTok{4}\NormalTok{, }\AttributeTok{format =} \StringTok{"d"}\NormalTok{),}
          \StringTok{"lambda"}\NormalTok{,}
          \FunctionTok{formatC}\NormalTok{(}
\NormalTok{            hList[[i]]}\SpecialCharTok{$}\NormalTok{l,}
            \AttributeTok{width =} \DecValTok{10}\NormalTok{,}
            \AttributeTok{digits =} \DecValTok{6}\NormalTok{,}
            \AttributeTok{format =} \StringTok{"f"}
\NormalTok{          ),}
          \StringTok{"stress"}\NormalTok{,}
          \FunctionTok{formatC}\NormalTok{(}
\NormalTok{            hList[[i]]}\SpecialCharTok{$}\NormalTok{s,}
            \AttributeTok{width =} \DecValTok{8}\NormalTok{,}
            \AttributeTok{digits =} \DecValTok{6}\NormalTok{,}
            \AttributeTok{format =} \StringTok{"f"}
\NormalTok{          ),}
          \StringTok{"penalty"}\NormalTok{,}
          \FunctionTok{formatC}\NormalTok{(}
\NormalTok{            hList[[i]]}\SpecialCharTok{$}\NormalTok{t,}
            \AttributeTok{width =} \DecValTok{8}\NormalTok{,}
            \AttributeTok{digits =} \DecValTok{6}\NormalTok{,}
            \AttributeTok{format =} \StringTok{"f"}
\NormalTok{          ),}
          \StringTok{"}\SpecialCharTok{\textbackslash{}n}\StringTok{"}
\NormalTok{        )}
\NormalTok{      \}}
      \ControlFlowTok{if}\NormalTok{ (hList[[i]]}\SpecialCharTok{$}\NormalTok{t }\SpecialCharTok{\textless{}}\NormalTok{ cut) \{}
\NormalTok{        mm }\OtherTok{\textless{}{-}}\NormalTok{ i}
        \ControlFlowTok{break}
\NormalTok{      \}}
\NormalTok{    \}}
    \FunctionTok{return}\NormalTok{(hList[}\DecValTok{1}\SpecialCharTok{:}\NormalTok{mm])}
\NormalTok{  \}}

\NormalTok{writeSelected }\OtherTok{\textless{}{-}} \ControlFlowTok{function}\NormalTok{ (hList, ind) \{}
\NormalTok{  m }\OtherTok{\textless{}{-}} \FunctionTok{length}\NormalTok{ (hList)}
\NormalTok{  n }\OtherTok{\textless{}{-}} \FunctionTok{length}\NormalTok{ (ind)}
\NormalTok{  mn }\OtherTok{\textless{}{-}} \FunctionTok{sort}\NormalTok{ (}\FunctionTok{union}\NormalTok{ (}\FunctionTok{union}\NormalTok{ (}\DecValTok{1}\SpecialCharTok{:}\DecValTok{3}\NormalTok{, ind), m }\SpecialCharTok{{-}}\NormalTok{ (}\DecValTok{2}\SpecialCharTok{:}\DecValTok{0}\NormalTok{)))}
  \ControlFlowTok{for}\NormalTok{ (i }\ControlFlowTok{in}\NormalTok{ mn) \{}
    \ControlFlowTok{if}\NormalTok{ (i }\SpecialCharTok{\textgreater{}}\NormalTok{ m) \{}
      \ControlFlowTok{next}
\NormalTok{    \}}
    \FunctionTok{cat}\NormalTok{(}
      \StringTok{"itel"}\NormalTok{,}
      \FunctionTok{formatC}\NormalTok{(hList[[i]]}\SpecialCharTok{$}\NormalTok{itel, }\AttributeTok{width =} \DecValTok{4}\NormalTok{, }\AttributeTok{format =} \StringTok{"d"}\NormalTok{),}
      \StringTok{"lambda"}\NormalTok{,}
      \FunctionTok{formatC}\NormalTok{(}
\NormalTok{        hList[[i]]}\SpecialCharTok{$}\NormalTok{l,}
        \AttributeTok{width =} \DecValTok{10}\NormalTok{,}
        \AttributeTok{digits =} \DecValTok{6}\NormalTok{,}
        \AttributeTok{format =} \StringTok{"f"}
\NormalTok{      ),}
      \StringTok{"stress"}\NormalTok{,}
      \FunctionTok{formatC}\NormalTok{(}
\NormalTok{        hList[[i]]}\SpecialCharTok{$}\NormalTok{s,}
        \AttributeTok{width =} \DecValTok{8}\NormalTok{,}
        \AttributeTok{digits =} \DecValTok{6}\NormalTok{,}
        \AttributeTok{format =} \StringTok{"f"}
\NormalTok{      ),}
      \StringTok{"penalty"}\NormalTok{,}
      \FunctionTok{formatC}\NormalTok{(}
\NormalTok{        hList[[i]]}\SpecialCharTok{$}\NormalTok{t,}
        \AttributeTok{width =} \DecValTok{8}\NormalTok{,}
        \AttributeTok{digits =} \DecValTok{6}\NormalTok{,}
        \AttributeTok{format =} \StringTok{"f"}
\NormalTok{      ),}
      \StringTok{"}\SpecialCharTok{\textbackslash{}n}\StringTok{"}
\NormalTok{    )}
\NormalTok{  \}}
\NormalTok{\}}
\end{Highlighting}
\end{Shaded}

\subsection{matchMe.R}\label{matchme.r}

\begin{Shaded}
\begin{Highlighting}[]
\NormalTok{matchMe }\OtherTok{\textless{}{-}} \ControlFlowTok{function}\NormalTok{ (x,}
                     \AttributeTok{itmax =} \DecValTok{100}\NormalTok{,}
                     \AttributeTok{eps =} \FloatTok{1e{-}10}\NormalTok{,}
                     \AttributeTok{verbose =} \ConstantTok{FALSE}\NormalTok{) \{}
\NormalTok{  m }\OtherTok{\textless{}{-}} \FunctionTok{length}\NormalTok{ (x)}
\NormalTok{  y }\OtherTok{\textless{}{-}} \FunctionTok{sumList}\NormalTok{ (x) }\SpecialCharTok{/}\NormalTok{ m}
\NormalTok{  itel }\OtherTok{\textless{}{-}} \DecValTok{1}
\NormalTok{  fold }\OtherTok{\textless{}{-}} \FunctionTok{sum}\NormalTok{ (}\FunctionTok{sapply}\NormalTok{ (x, }\ControlFlowTok{function}\NormalTok{ (z)}
\NormalTok{    (z }\SpecialCharTok{{-}}\NormalTok{ y) }\SpecialCharTok{\^{}} \DecValTok{2}\NormalTok{))}
  \ControlFlowTok{repeat}\NormalTok{ \{}
    \ControlFlowTok{for}\NormalTok{ (j }\ControlFlowTok{in} \DecValTok{1}\SpecialCharTok{:}\NormalTok{m) \{}
\NormalTok{      u }\OtherTok{\textless{}{-}} \FunctionTok{crossprod}\NormalTok{ (x[[j]], y)}
\NormalTok{      s }\OtherTok{\textless{}{-}} \FunctionTok{svd}\NormalTok{ (u)}
\NormalTok{      r }\OtherTok{\textless{}{-}} \FunctionTok{tcrossprod}\NormalTok{ (s}\SpecialCharTok{$}\NormalTok{u, s}\SpecialCharTok{$}\NormalTok{v)}
\NormalTok{      x[[j]] }\OtherTok{\textless{}{-}}\NormalTok{ x[[j]] }\SpecialCharTok{\%*\%}\NormalTok{ r}
\NormalTok{    \}}
\NormalTok{    y }\OtherTok{\textless{}{-}} \FunctionTok{sumList}\NormalTok{ (x) }\SpecialCharTok{/}\NormalTok{ m}
\NormalTok{    fnew }\OtherTok{\textless{}{-}} \FunctionTok{sum}\NormalTok{ (}\FunctionTok{sapply}\NormalTok{ (x, }\ControlFlowTok{function}\NormalTok{ (z)}
\NormalTok{      (z }\SpecialCharTok{{-}}\NormalTok{ y) }\SpecialCharTok{\^{}} \DecValTok{2}\NormalTok{))}
    \ControlFlowTok{if}\NormalTok{ (verbose) \{}
      
\NormalTok{    \}}
    \ControlFlowTok{if}\NormalTok{ (((fold }\SpecialCharTok{{-}}\NormalTok{ fnew) }\SpecialCharTok{\textless{}}\NormalTok{ eps) }\SpecialCharTok{||}\NormalTok{ (itel }\SpecialCharTok{==}\NormalTok{ itmax))}
      \ControlFlowTok{break}
\NormalTok{    itel }\OtherTok{\textless{}{-}}\NormalTok{ itel }\SpecialCharTok{+} \DecValTok{1}
\NormalTok{    fold }\OtherTok{\textless{}{-}}\NormalTok{ fnew}
\NormalTok{  \}}
  \FunctionTok{return}\NormalTok{ (x)}
\NormalTok{\}}

\NormalTok{sumList }\OtherTok{\textless{}{-}} \ControlFlowTok{function}\NormalTok{ (x) \{}
\NormalTok{  m }\OtherTok{\textless{}{-}} \FunctionTok{length}\NormalTok{ (x)}
\NormalTok{  y }\OtherTok{\textless{}{-}}\NormalTok{ x[[}\DecValTok{1}\NormalTok{]]}
  \ControlFlowTok{for}\NormalTok{ (j }\ControlFlowTok{in} \DecValTok{2}\SpecialCharTok{:}\NormalTok{m) \{}
\NormalTok{    y }\OtherTok{\textless{}{-}}\NormalTok{ y }\SpecialCharTok{+}\NormalTok{ x[[j]]}
\NormalTok{  \}}
  \FunctionTok{return}\NormalTok{ (y)}
\NormalTok{\}}
\end{Highlighting}
\end{Shaded}

\subsection{plotMe.R}\label{plotme.r}

\begin{Shaded}
\begin{Highlighting}[]
\NormalTok{plotMe2 }\OtherTok{\textless{}{-}} \ControlFlowTok{function}\NormalTok{(hList, labels, }\AttributeTok{s =} \DecValTok{1}\NormalTok{, }\AttributeTok{t =} \DecValTok{2}\NormalTok{) \{}
\NormalTok{  n }\OtherTok{\textless{}{-}} \FunctionTok{nrow}\NormalTok{(hList[[}\DecValTok{1}\NormalTok{]]}\SpecialCharTok{$}\NormalTok{x)}
\NormalTok{  m }\OtherTok{\textless{}{-}} \FunctionTok{length}\NormalTok{ (hList)}
  \FunctionTok{par}\NormalTok{(}\AttributeTok{pty =} \StringTok{"s"}\NormalTok{)}
\NormalTok{  hMatch }\OtherTok{\textless{}{-}} \FunctionTok{matchMe}\NormalTok{ (}\FunctionTok{lapply}\NormalTok{ (hList, }\ControlFlowTok{function}\NormalTok{(r)}
\NormalTok{    r}\SpecialCharTok{$}\NormalTok{x))}
\NormalTok{  hMat }\OtherTok{\textless{}{-}} \FunctionTok{matrix}\NormalTok{ (}\DecValTok{0}\NormalTok{, }\DecValTok{0}\NormalTok{, }\DecValTok{2}\NormalTok{)}
  \ControlFlowTok{for}\NormalTok{ (j }\ControlFlowTok{in} \DecValTok{1}\SpecialCharTok{:}\NormalTok{m) \{}
\NormalTok{    hMat }\OtherTok{\textless{}{-}} \FunctionTok{rbind}\NormalTok{(hMat, hMatch[[j]][, }\FunctionTok{c}\NormalTok{(s, t)])}
\NormalTok{  \}}
  \FunctionTok{plot}\NormalTok{(hMat,}
       \AttributeTok{xlab =} \StringTok{"dim 1"}\NormalTok{,}
       \AttributeTok{ylab =} \StringTok{"dim 2"}\NormalTok{,}
       \AttributeTok{col =} \FunctionTok{c}\NormalTok{(}\FunctionTok{rep}\NormalTok{(}\StringTok{"RED"}\NormalTok{, n}\SpecialCharTok{*}\NormalTok{(m}\DecValTok{{-}1}\NormalTok{)), }\FunctionTok{rep}\NormalTok{(}\StringTok{"BLUE"}\NormalTok{, n)),}
       \AttributeTok{cex =} \FunctionTok{c}\NormalTok{(}\FunctionTok{rep}\NormalTok{(}\DecValTok{1}\NormalTok{, n}\SpecialCharTok{*}\NormalTok{(m}\DecValTok{{-}1}\NormalTok{)), }\FunctionTok{rep}\NormalTok{(}\DecValTok{2}\NormalTok{, n)))}
  \ControlFlowTok{for}\NormalTok{ (i }\ControlFlowTok{in} \DecValTok{1}\SpecialCharTok{:}\NormalTok{n) \{}
\NormalTok{    hLine }\OtherTok{\textless{}{-}} \FunctionTok{matrix}\NormalTok{ (}\DecValTok{0}\NormalTok{, }\DecValTok{0}\NormalTok{, }\DecValTok{2}\NormalTok{)}
    \ControlFlowTok{for}\NormalTok{ (j }\ControlFlowTok{in} \DecValTok{1}\SpecialCharTok{:}\NormalTok{m) \{}
\NormalTok{      hLine }\OtherTok{\textless{}{-}} \FunctionTok{rbind}\NormalTok{ (hLine, hMatch[[j]][i, }\FunctionTok{c}\NormalTok{(s, t)])}
\NormalTok{    \}}
    \FunctionTok{lines}\NormalTok{(hLine)}
\NormalTok{  \}}
  \FunctionTok{text}\NormalTok{(hMatch[[m]], labels, }\AttributeTok{cex =}\NormalTok{ .}\DecValTok{75}\NormalTok{)}
\NormalTok{\}}

\NormalTok{plotMe1 }\OtherTok{\textless{}{-}} \ControlFlowTok{function}\NormalTok{(hList, labels) \{}
\NormalTok{  n }\OtherTok{\textless{}{-}} \FunctionTok{length}\NormalTok{ (hList[[}\DecValTok{1}\NormalTok{]]}\SpecialCharTok{$}\NormalTok{x)}
\NormalTok{  m }\OtherTok{\textless{}{-}} \FunctionTok{length}\NormalTok{ (hList)}
\NormalTok{  blow }\OtherTok{\textless{}{-}} \ControlFlowTok{function}\NormalTok{ (x) \{}
\NormalTok{    n }\OtherTok{\textless{}{-}} \FunctionTok{length}\NormalTok{ (x)}
    \FunctionTok{return}\NormalTok{ (}\FunctionTok{matrix}\NormalTok{ (}\FunctionTok{c}\NormalTok{(}\DecValTok{1}\SpecialCharTok{:}\NormalTok{n, x), n, }\DecValTok{2}\NormalTok{))}
\NormalTok{  \}}
\NormalTok{  hMat }\OtherTok{\textless{}{-}} \FunctionTok{matrix}\NormalTok{ (}\DecValTok{0}\NormalTok{, }\DecValTok{0}\NormalTok{, }\DecValTok{2}\NormalTok{)}
  \ControlFlowTok{for}\NormalTok{ (j }\ControlFlowTok{in} \DecValTok{1}\SpecialCharTok{:}\NormalTok{m) \{}
\NormalTok{    hMat }\OtherTok{\textless{}{-}} \FunctionTok{rbind}\NormalTok{(hMat, }\FunctionTok{blow}\NormalTok{(hList[[j]]}\SpecialCharTok{$}\NormalTok{x))}
\NormalTok{  \}}
  \FunctionTok{plot}\NormalTok{(hMat,}
       \AttributeTok{xlab =} \StringTok{"index"}\NormalTok{,}
       \AttributeTok{ylab =} \StringTok{"x"}\NormalTok{,}
       \AttributeTok{col =} \FunctionTok{c}\NormalTok{(}\FunctionTok{rep}\NormalTok{(}\StringTok{"RED"}\NormalTok{, n}\SpecialCharTok{*}\NormalTok{(m}\DecValTok{{-}1}\NormalTok{)), }\FunctionTok{rep}\NormalTok{(}\StringTok{"BLUE"}\NormalTok{, n)),}
       \AttributeTok{cex =} \FunctionTok{c}\NormalTok{(}\FunctionTok{rep}\NormalTok{(}\DecValTok{1}\NormalTok{, n}\SpecialCharTok{*}\NormalTok{(m}\DecValTok{{-}1}\NormalTok{)), }\FunctionTok{rep}\NormalTok{(}\DecValTok{2}\NormalTok{, n)))}
  \ControlFlowTok{for}\NormalTok{ (i }\ControlFlowTok{in} \DecValTok{1}\SpecialCharTok{:}\NormalTok{n) \{}
\NormalTok{    hLine }\OtherTok{\textless{}{-}} \FunctionTok{matrix}\NormalTok{ (}\DecValTok{0}\NormalTok{, }\DecValTok{0}\NormalTok{, }\DecValTok{2}\NormalTok{)}
    \ControlFlowTok{for}\NormalTok{ (j }\ControlFlowTok{in} \DecValTok{1}\SpecialCharTok{:}\NormalTok{m) \{}
\NormalTok{      hLine }\OtherTok{\textless{}{-}} \FunctionTok{rbind}\NormalTok{ (hLine, }\FunctionTok{blow}\NormalTok{(hList[[j]]}\SpecialCharTok{$}\NormalTok{x)[i, ])}
      \FunctionTok{lines}\NormalTok{(hLine)}
\NormalTok{    \}}
\NormalTok{  \}}
  \FunctionTok{text}\NormalTok{(}\FunctionTok{blow}\NormalTok{(hList[[m]]}\SpecialCharTok{$}\NormalTok{x), labels, }\AttributeTok{cex =} \FloatTok{1.00}\NormalTok{)}
  \ControlFlowTok{for}\NormalTok{ (i }\ControlFlowTok{in} \DecValTok{1}\SpecialCharTok{:}\NormalTok{n) \{}
    \FunctionTok{abline}\NormalTok{(}\AttributeTok{h =}\NormalTok{ hList[[m]]}\SpecialCharTok{$}\NormalTok{x[i])}
\NormalTok{  \}}
\NormalTok{\}}
\end{Highlighting}
\end{Shaded}

\subsection{checkUni.R}\label{checkuni.r}

\begin{Shaded}
\begin{Highlighting}[]
\NormalTok{checkUni }\OtherTok{\textless{}{-}} \ControlFlowTok{function}\NormalTok{ (w, delta, x) \{}
\NormalTok{  x }\OtherTok{\textless{}{-}} \FunctionTok{drop}\NormalTok{ (x)}
\NormalTok{  n }\OtherTok{\textless{}{-}} \FunctionTok{length}\NormalTok{ (x)}
\NormalTok{  vinv }\OtherTok{\textless{}{-}} \FunctionTok{solve}\NormalTok{ (}\FunctionTok{smacofVmat}\NormalTok{ (w) }\SpecialCharTok{+}\NormalTok{ (}\DecValTok{1} \SpecialCharTok{/}\NormalTok{ n)) }\SpecialCharTok{{-}}\NormalTok{ (}\DecValTok{1} \SpecialCharTok{/}\NormalTok{ n)}
  \FunctionTok{return}\NormalTok{ (}\FunctionTok{drop}\NormalTok{ (vinv }\SpecialCharTok{\%*\%} \FunctionTok{rowSums}\NormalTok{ (w }\SpecialCharTok{*}\NormalTok{ delta }\SpecialCharTok{*} \FunctionTok{sign}\NormalTok{ (}\FunctionTok{outer}\NormalTok{ (x, x, }\StringTok{"{-}"}\NormalTok{)))))}
\NormalTok{\}}
\end{Highlighting}
\end{Shaded}

\section*{References}\label{references}
\addcontentsline{toc}{section}{References}

\phantomsection\label{refs}

\end{document}